\documentclass[a4paper,fleqn,usenatbib]{mnras}

\usepackage{newtxtext,newtxmath}
\usepackage[normalem]{ulem}

\usepackage[T1]{fontenc}
\usepackage{ae,aecompl}

\usepackage{graphicx}	% Including figure files
\usepackage{amsmath}	% Advanced maths commands
\usepackage{xspace} 
\usepackage{booktabs}

\newcommand*\blu[0]{\hbox{$\text{bl}_{\text{u}}$}\xspace}
\newcommand*\blo[0]{\hbox{$\text{bl}_{\text{o}}$}\xspace}
\DeclareMathOperator{\med}{med}

\newcommand*{\orcidlink}[1]{%
	\href{https://orcid.org/#1}{\includegraphics{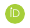}}%
}
\title[B/PS bulges and barlenses]{B/PS bulges and barlenses from a kinematic viewpoint. I}

\author[Daria Zakharova et al.]{Daria Zakharova$^{1,2}$\orcidlink{0009-0001-1809-4821}\thanks{E-mail: dzakharovaa@gmail.com}, Iliya S. Tikhonenko$^{3}$, Natalia Ya. Sotnikova$^{4}$, and Anton A. Smirnov$^{5}$\\
$^{1}$Dipartimento di Fisica e Astronomia "Galileo Galilei", Universita' degli studi di Padova, Vicolo dell'Osservatorio, 3, I-35122, Padova, Italy\\
$^{2}$INAF - Osservatorio astronomico di Padova, Vicolo dell'Osservatorio, 5, I-35122, Padova, Italy\\
$^{3}$Max-Planck-Institut für extraterrestrische Physik, Gießenbachstraße, D-85748 Garching, Germany\\
$^{4}$St. Petersburg State University,
Universitetskij pr.~28, 198504 St. Petersburg, Stary Peterhof, Russia\\
$^{5}$Central (Pulkovo) Astronomical Observatory of RAS, Pulkovskoye Chaussee 65/1, 196140 St. Petersburg, Russia\\
}
\date{Accepted XXX. Received YYY; in original form ZZZ}

\pubyear{2023}

\begin{document}
\label{firstpage}
\pagerange{\pageref{firstpage}--\pageref{lastpage}}
\maketitle

\begin{abstract}
A significant part of barred disc galaxies exhibits boxy/peanut-shaped structures (B/PS bulges) at high inclinations. 
Another structure also associated with the bar is a barlens, often observed in galaxies in a position close to face-on.
At this viewing angle, special kinematic tests are required to detect a 3D extension of the bars in the vertical direction (B/PS bulges). We use four pure $N$-body models of galaxies with B/PS bulges, which have different bar morphology from bars with barlenses to the so-called face-on peanut bars. We analyse the kinematics of our models to establish how the structural features of B/PS bulges manifest themselves in the kinematics for galaxies at intermediate inclinations and whether these features are related to the barlenses. We apply the dissection of the bar into different orbital groups to determine which of them are responsible for the features of the LOSVD (line-of-sight velocity distribution), i.e., for the deep minima of the $h_4$ parameter along the major axis of the bar. As a result, we claim that for our models at the face-on position, the kinematic signatures of a `peanut' indeed track the vertical density distribution features. We conclude that orbits responsible for such kinematic signatures differ from model to model. We pay special attention to the barlens model. We show that orbits assembled into barlens are not responsible for the kinematic signatures of B/PS bulges. The results presented in this work are applicable to the interpretation of IFU observations of real galaxies.
\end{abstract}
%%%%%%%%%%%%%%%%%%%%%%%%%%%%%%%%%%%%%%%%%%%%%%%%%%
%%%%%%%%%%%%%%%%%%%%%%%%%%%%%%%%%%%%%%%%%%%%%%%%%%


%%%%%%%%%%%%%%%%%%%%%%%%%%%%%%%%%%%%%%%%%%%%%%%%%%
%%%%%%%%%%%%%%%%% BODY OF PAPER %%%%%%%%%%%%%%%%%%

%%%%%%%%%%%%%%%%%%%%%%%%%%%%%%%%%%%%%%%%%%%%%%%%%%
%%%%%%%%%%%%%%%%%%%%%%%%%%%%%%%%%%%%%%%%%%%%%%%%%%
\section{Introduction}
\label{sec:intro}
% 1.
%%%%%%%%%%%%%%%%%%%%%%%%%%%%%%%%%%%%%%%%%%%%%%%%%%
%%%%%%%%%%%%%%%%%%%%%%%%%%%%%%%%%%%%%%%%%%%%%%%%%%
The barlens component has been introduced only recently in morphological classifications of nearly face-on galaxies. It is described as a new type of lens embedded into the bar and covering about half its length \citep{Laurikainen_etal2011}. Its outline is circular-like, while its azimuth-averaged profile is exponential, which distinguishes barlenses from classical bulges \citep{Laurikainen_Salo2017,Athanassoula_etal2015}. The barlenses are typically four times larger than nuclear discs and rings and are also distinct from standard lenses \citep{Kormendy1979} because they are smaller than bars \citep{Laurikainen_etal2018,Athanassoula_etal2015}.
\par
The so-called B/PS (boxy/peanut-shaped) bulges are also associated with the bars and stand out well against the edge-on stellar disc background \citep{Lutticke_etal2000a}. B/PS bulges are the thickest inner parts of the bars, which have grown vertically as the bars have evolved \citep{Combes_Sanders1981,Athanassoula2016,Laurikainen_Salo2016}.
Although B/PS structures in $N$-body simulations inevitably form as parts of bars \citep{Combes_Sanders1981}, the relationship between B/PS structures and bars is not obvious in observations due to different viewing conditions. This connection is established through the statistics of large samples of galaxies.
\citet{Lutticke_etal2000a} (see also \citealp{Li_etal2017,Erwin_Debattista2017}) found that up to 45\% of all S0-Sd galaxies have B/PS structures, which is close to the fraction of barred galaxies ($\sim$ 50\%-60\%, \citealp{Marinova_Jogee2007}). \citet{Erwin_Debattista2013,Erwin_Debattista2017} found that B/PS features extend from 1/3 to 2/3 of the bar radius. In addition to the statistical arguments, kinematic arguments also work: in edge-on galaxies with B/PS structures, traces of a bar can be found from the kinematic characteristics of elongated orbits \citep{Bureau_Athanassoula2005,Iannuzzi_Athanassoula2015}. 
\par
Thus, two structural features of galaxies, B/PS bulges and barlenses, are closely associated with bars. 
B/PS bulges stand out when the galaxy is viewed edge-on, and barlenses are well identified visually when the galaxy positioned almost face-on. That is why the idea has recently spread that `barlens' and `box/peanut' are the same stellar structure, namely the inner part of the bar, but seen at different viewing angles \citep{Laurikainen_etal2014,Athanassoula2016}. 
The identification of the two structures was based on several arguments, mostly indirect. 
The parent galaxies of barlenses and B/PS structures have a similar total stellar mass distribution \citep{Laurikainen_etal2014}.
Relative to the bar, barlenses have similar sizes \citep{Athanassoula_etal2015} as obtained for the B/PS bulges by \citet{Lutticke_etal2000b} and \citet{Erwin_Debattista2013}. However, the main evidence came from analysing the results of $N$-body simulations. By viewing their snapshots from many different angles, \citet{Athanassoula_etal2015} concluded that barlenses are simply the vertically thick part of the bar, i.e., the barlenses in the edge-on position appear as B/PS bulges. 
Stronger evidence comes from an analysis of the kinematics of galaxies with barlenses. First, \citet{Mendez-Abreu_etal2008} found kinematic evidence of the B/PS bulge for the galaxy NGC~98 classified as having a barlens \citep{Laurikainen_Salo2017}. Then, for NGC 1640, which classified as a galaxy with a barlens \citep{Laurikainen_Salo2017}, \citet{Mendez-Abreu_etal2014} showed a kinematic hint of the presence of a B/PS bulge in the central regions of the bar. Finally, the TIMER team \citep{Gadotti_etal2020} claim to have found in the barlens region of several nearly face-on galaxies kinematic signatures of both a bar and a box/peanut, as expected from numerical simulations. 
But for almost face-on galaxies it is not yet clear which structural component of the bar (the bar itself or the barlens) is associated with the discovered kinematic signatures of a `peanut'. That is why even the kinematic data needs a more accurate interpretation.
\par
This question can be solved using the results of the orbital analysis of numerical models. Recently \citet{Tikhonenko_etal2021} have dissected a bar in one typical $N$-body model with a barlens into separate orbital components supported by completely different types of orbits. This work shows that the part of the barlens with rounded isophotes in the face-on view is a rather flat structure in the vertical direction without any significant off-centre protrusions and signatures of a peanut. 
It was also concluded that the edge-on B/PS bulge is a barlens-independent component of the bar and is supported by various types of orbits. One can not separate the face-on image of the galaxy into individual orbital components and find kinematic parameters for these components, but by using the results of $N$-body simulations and orbital analysis, one can understand which orbital subsystem contributes to the kinematic features of a peanut.
\par
Four parameters are often used in kinematic diagnostics:
the mean velocity $\overline V$, the velocity dispersion $\sigma$, and $h_3$, $h_4$ parameters, which measure the deviations of the LOSVD (line-of-sight velocity distribution) from a Gaussian. 
The last two parameters are the corresponding coefficients that arise when the velocity profile is expanded by orthogonal Gauss-Hermite functions.
The parameter $h_3$ measures asymmetric deviations from the Gaussian, while the parameter $h_4$ measures symmetric ones. These deviations occur when different velocity components on the line of sight do not contribute equally to the LOSVD \citep{vanderMarel_Franx1993,Gerhard1993}.
\par
\citet{Debattista_etal2005} used kinematic diagnostic to analyse the LOSVD of barred face-on galaxies. In this case, there is a single velocity component, a vertical one ($v_z$).
\citet{Debattista_etal2005} noted that the velocity distribution (VD) for the vertical component $f(v_z)$ may differ from the Gaussian and lead to negative values of $h_4$ if the density distribution in the vertical direction is flat-topped, i.e., not so strongly concentrated towards the mid-plane compared to the surrounding disc. The flat-topped volume density distribution in the vertical direction is expressed in terms of negative values of the fourth coefficient of Gauss-Hermite expansions $d4$~(an analogue of $h_4$).
It opens up the possibility of confirming the relationship between bars~(structures that are well identified in a nearly face-on view) and B/PS bulges, which are seen predominantly edge-on. \citet{Debattista_etal2005} have noticed that negative values of $h_4$ along the major axis of the bar at inclinations less than $30^\circ$ indicate the presence of the peanut if they form two deep minima on both sides of the centre of the galaxy. Two shallow minima of $h_4$ along the major axis of the bar indicate not a peanut, but a contamination of the LOSVD by other velocity components, in addition to $v_z$.
\par
The origin of the negative $h_4$ values is still not clearly understood while being particularly important in the light of recent high resolution integral field unit (IFU) spectroscopy data (for instance, the TIMER project, \citealp{Gadotti_etal2019}). The accurate interpretation of these features (negative $h_4$ values)
%from the dynamical viewpoint 
would require a set of synthetic 2D maps of the common kinematic parameters ($\overline V$, $\sigma$ and $h_3$, $h_4$ parameters) based on a wide grid of numerical models viewed from different inclinations and bar major axis viewing angles.
\par
We intend to explore model kinematic maps for intermediate inclinations, thus extending the data available from
\citet{Debattista_etal2005} for a face-on view and from \citet{Iannuzzi_Athanassoula2015} for mainly high inclined models. In addition, since quite a lot of galaxies with barlenses fall into the sample by \citet{Gadotti_etal2020}, we would like to conduct a comparative analysis of 2D maps for models with gradually changing bar morphology from models with barlenses to models without features in the face-on structure of the bar. We also intend to collect the analysed model data into data cubes with an appropriate spectral resolution for the most plausible comparison with observational data cubes. Such maps will be useful both for interpreting existing and future observational data. Since 2D kinematic maps for inclined galaxies can differ from face-on maps, it is necessary to extend the analysis of the LOSVD distortions due to i) density distribution features, ii) contamination of the LOSVD by various velocity components. This will help to better understand the relationship of kinematic parameters and B/PS bulges in cases when the B/PS bulges themselves are not visible. Finally, having the results of an orbital analysis for a number of numerical models, we have a unique opportunity to study the influence of different orbital groups on the kinematic picture and to understand which orbital groups~(assembled into a barlens or into the bar itself) raise distortions into the LOSVD and create peanut signatures when the galaxy is viewed face-on.
\par
In this article, we focus only on the features of the LOSVD, which are measured by the parameter $h_4$. We leave the joint analysis of the parameters $h_3$ and $h_4$ for future works.
\par
The paper is organised as follows.
In Section~\ref{sec:data}, we describe our $N$-body models as well as our dissection of the models into various orbital groups assembled into different sub-strictures of the bar, explain how we prepare data cubes for all models and for a set of inclinations and position angles of a bar and discuss how we extract the main 
kinematic parameters from LOSVDs in each pixel. 
In Section~\ref{sec:LOSVD}, we present our results, compare 2D kinematic maps for models with a barlens and with an ordinary bar at face-on view and other positions of a bar. In Section~\ref{sec:LOSVD_vel_components}, we differentiate the effects of a flat-topped vertical density distribution and effects of inclinations associated with contamination of the LOSVD by velocity components other than $v_z$.
In Section~\ref{sec:orbital_groups}, we focus on the contributions of different orbital groups into features of the LOSVD. In Section~\ref{sec:discussion} we discuss the astrophysical applications of our results. In Section~\ref{sec:conclusions}, we give our conclusions.

%%%%%%%%%%%%%%%%%%%%%%%%%%%%%%%%%%%%%%%%%%%%%%%%%%
%%%%%%%%%%%%%%%%%%%%%%%%%%%%%%%%%%%%%%%%%%%%%%%%%%
\section{Data preparation}
\label{sec:data}
% 2.
%%%%%%%%%%%%%%%%%%%%%%%%%%%%%%%%%%%%%%%%%%%%%%%%%%
%%%%%%%%%%%%%%%%%%%%%%%%%%%%%%%%%%%%%%%%%%%%%%%%%%

%%%%%%%%%%%%%%%%%%%%%%%%%%%%%%%%%%%%%%%%%%%%%%%%%%
%%%%%%%%%%%%%%%%%%%%%%%%%%%%%%%%%%%%%%%%%%%%%%%%%%
\subsection{N-body models}
\label{sec:nbody}
% 2.1.
%%%%%%%%%%%%%%%%%%%%%%%%%%%%%%%%%%%%%%%%%%%%%%%%%%
%%%%%%%%%%%%%%%%%%%%%%%%%%%%%%%%%%%%%%%%%%%%%%%%%%
To study the kinematic properties of B/PS bulges, we use a set of self-consistent pure $N$-body (i.e., without the gas component) models from \citet{Smirnov_etal2021}.
In the following description and most of the illustrations, we adopt the unit system where 1 length unit is equal to $3.5$ kpc, $G=1$, and unit mass is equal 
to $5\times10^{10}M_{\odot}$. Assuming these values, a time unit would be equal to $13.8$~Myr. 
\par
The initial conditions for each model were prepared with \texttt{mkgalaxy} script \citep{McMillan_Dehnen_2007}. At the start of simulations, all models have a stellar disc embedded into a self-consistent (``live'') dark halo.  
\par 
The stellar disc of each model has an exponential radial profile and is 
isothermal in the vertical direction:
%%%%%%%%%%%%%%%%%%%%%%%%%%%%%%%%%%%%%%%%%%%%%%%%%%
%%%%%%%%%%%%%%%%%%%%%%%%%%%%%%%%%%%%%%%%%%%%%%%%%%
\begin{equation}
\rho_\mathrm{d}(R,z)=\frac{M_\mathrm{d}}{4\pi R_\mathrm{d}^2 z_\mathrm{d}} \cdot \exp(-R/R_\mathrm{d}) \cdot \mathrm{sech}^2(z/z_\mathrm{d}) \,,
\label{eq:rho_disk} 
\end{equation}
%%%%%%%%%%%%%%%%%%%%%%%%%%%%%%%%%%%%%%%%%%%%%%%%%%
%%%%%%%%%%%%%%%%%%%%%%%%%%%%%%%%%%%%%%%%%%%%%%%%%%
where $R$ is the cylindrical radius, $M_\mathrm{d}=1$ is the total mass of the disc, $R_\mathrm{d}=1$ and $z_\mathrm{d}=0.05$ are scale lengths in radial and vertical directions, respectively.
\par
The halos are spherical and have Navarro-Frenk-White-like~\citep{NFW} profile with dark matter mass for each model equal to $M_\mathrm{h}(r < 4R_\mathrm{d})/M_\mathrm{d} \approx 1.5$.
\par
\begin{figure}
\centering
\includegraphics[width=1\linewidth]{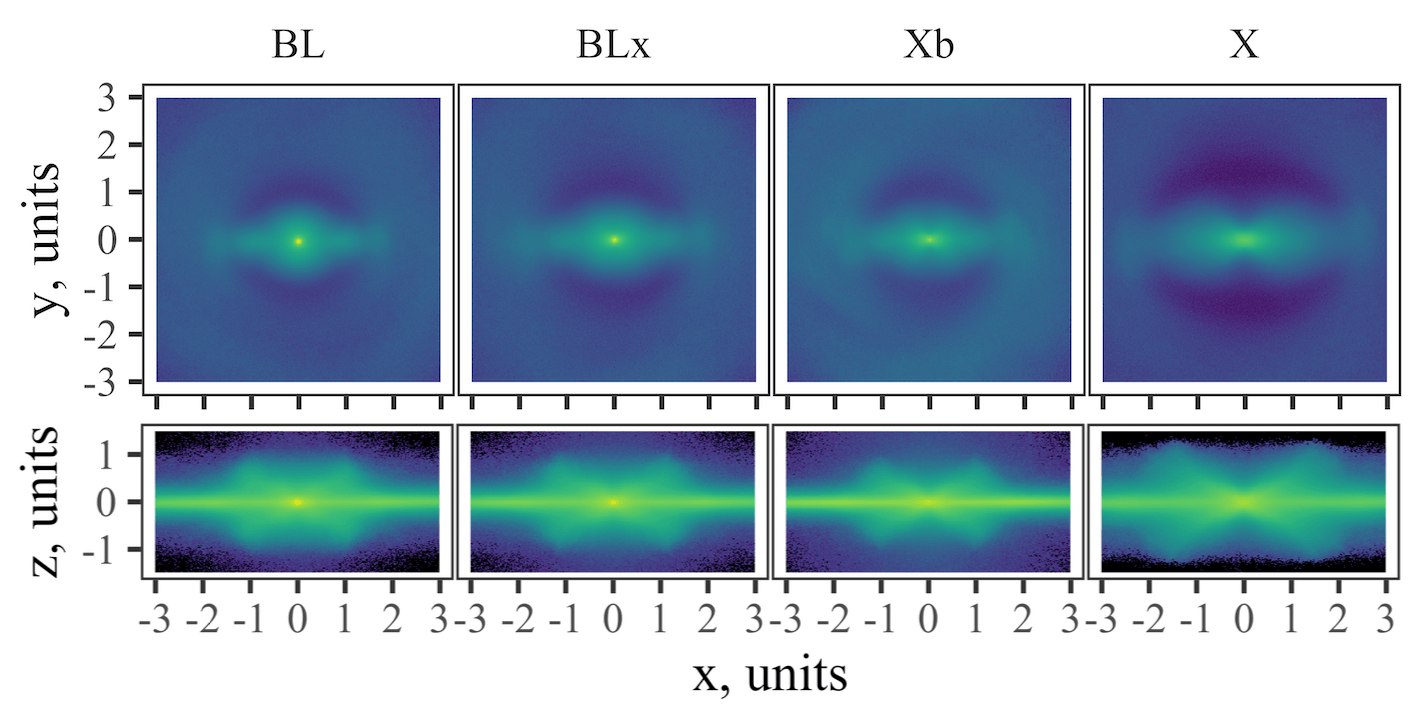}
\caption{The set of models (BL, BLx, Xb, and X) from \citet{Smirnov_etal2021} used in this work. The \textit{top} row shows face-on views of the corresponding model in $(x,y)=[-3; 3]\times [-3;3]$ area, while the \textit{bottom} row displays  edge-on views in $(x,z)=[-3; 3]\times[-1.5; 1.5]$. 
}%
\label{fig:models_view}
\end{figure}

For three models, a spherical Hernquist bulge \citep{Hernquist1990} in the central area was added. All models in the course of their evolution develop bars with B/PS bulges and X-structures (the brightest part of the B/PS bulges) visible at the edge-on position. We distinguish models by their nicknames according to the final morphology of their bars: BL~(BarLens), BLx~(BarLens + slight traces of face-on X-shaped bar), Xb~(face-on X-shaped bar + bulge), X~(face-on X-shaped bar). BL and BLx models are the galaxies with a barlens in a face-on position~(first and second columns in Fig.~\ref{fig:models_view}). X and Xb models represent galaxies with X-shaped (or peanut-shaped) bars in a face-on position~(third and fourth columns in Fig.~\ref{fig:models_view}). Such a morphology difference is due to the presence/absence of the bulge. The bulges in models differ in the total mass $M_\mathrm{b}$ and radial scale lengths $r_\mathrm{b}$ (Table~\ref{tab:modelparams}). \footnote{\textcolor{black}{One should bear in mind  that galaxies which show X-shape at low inclinations are quite rare.}}
\par
Each subsystem was represented by a specific number of ``star''-particles: 
\textcolor{black}{$4\times10^6$}
%\ \textit{kk}$ 
for the disc, \textcolor{black}{$4.5\times10^6$}
%\,\textit{kk}$ 
for the halo. 
The total number of bulge particles is determined according to the value of the ratio $M_\mathrm{b}/M_\mathrm{d}$, with a reference point for the Xb model, where the number is equal to \textcolor{black}{$0.8\times10^6$}. 
%\textit{kk}. 
\par
The disc radial dispersion profile was exponential with $\sigma_R=\sigma_0 \cdot \exp(-R/2 R_\mathrm{d})$, where $\sigma_0$ is a normalisation constant. It is inferred from the Toomre parameter value: $Q(2\,R_\mathrm{d})=Q_0$ with $Q_0$ set to 1.2. Thus, we consider a marginally stable stellar disc.
The evolution of all models was traced by \texttt{gyfaclON} integrator \citep{Dehnen2002} from the NEMO suite \citep{Teuben_1995} for approximately 8 Gyr with 1.7 Myr time step.  
Fig.~\ref{fig:models_view} shows snapshots for our models for $t=450$ ($\approx 6.2$~Gyr). 
\par
The key difference between the models is their initial bulge concentration, which results in different gradient of the rotation curve in the centre.
For example, the BL model has the steepest rotation curve in the central regions~(see \citealp{Smirnov_etal2021}, figure~1). The relation between the slope of the rotation curve in the inner parts of the galaxy and different face-on bar morphology was first noted by \citep{Salo_Laurikainen2017}. Our BL model does fall into the range of parameters for which the model develops a clear barlens. Thus, the chosen setup of models yields a gradual progression in face-on bar morphology from a bar with a barlens to a face-on peanut, which makes this model set optimal for understanding the nature of the barlens and its connection with a B/PS bulge: either in terms of its vertical structure (done in \citealt{Tikhonenko_etal2021}) or its kinematics (this work). 

%%%%%%%%%%%%%%%%%%%%%%%%%%%%%%%%%%%%%%%%%%%%%%%%%%
%%%%%%%%%%%%%%%%%%%%%%%%%%%%%%%%%%%%%%%%%%%%%%%%%%
% Fig.~1.
%%%%%%%%%%%%%%%%%%%%%%%%%%%%%%%%%%%%%%%%%%%%%%%%%%
%%%%%%%%%%%%%%%%%%%%%%%%%%%%%%%%%%%%%%%%%%%%%%%%%%

%%%%%%%%%%%%%%%%%%%%%%%%%%%%%%%%%%%%%%%%%%%%%%%%%%
%%%%%%%%%%%%%%%%%%%%%%%%%%%%%%%%%%%%%%%%%%%%%%%%%%
\begin{table}
\centering
\begin{tabular}{rcccc}
\toprule
model & X & Xb & BLx & BL \\ 
$M_\mathrm{b}$ & $0$ & $0.2$ & $0.1$ & $0.1$ \\
$r_\mathrm{b}$ & - & $0.2$ & $0.1$ & $0.05$ \\
\bottomrule
\end{tabular}
\caption{The properties of classical bulges in our models, see text for the adopted unit system description.}
\label{tab:modelparams}
\end{table}
%%%%%%%%%%%%%%%%%%%%%%%%%%%%%%%%%%%%%%%%%%%%%%%%%%
%%%%%%%%%%%%%%%%%%%%%%%%%%%%%%%%%%%%%%%%%%%%%%%%%%

%%%%%%%%%%%%%%%%%%%%%%%%%%%%%%%%%%%%%%%%%%%%%%%%%%
%%%%%%%%%%%%%%%%%%%%%%%%%%%%%%%%%%%%%%%%%%%%%%%%%%
\subsection{Orbital groups}
\label{sec:families}
% 2.2.
%%%%%%%%%%%%%%%%%%%%%%%%%%%%%%%%%%%%%%%%%%%%%%%%%%
%%%%%%%%%%%%%%%%%%%%%%%%%%%%%%%%%%%%%%%%%%%%%%%%%%
In \citet{Parul_etal2020} and \citet{Smirnov_etal2021}, the entire orbital structure of several self-consistent $N$-body models with bars were analysed using  Fourier analysis of spatial coordinates of each star-particle. Four models presented in Section~\ref{sec:nbody} fall into the grid of models for which orbital analysis was performed. Thus, we have unique data for studying the kinematics of individual orbital groups and searching for kinematic signatures of a peanut on face-on snapshots that we can attribute to one or another orbital group.
\par
The conducted orbital analysis is based on spectral dynamics methods \citep{Binney_Spergel1982}. For all orbits in the disc, the Cartesian frequencies ($f_x$, $f_y$, and $f_z$) in the rotating coordinate frame with $x$-axis aligned with the bar major axis\footnote{\textcolor{black}{To identify the bar major axis we diagonalise the moment-of-inertia tensor of a subset of particles ($R < 0.8$, $|z| < 0.3$) and find principal axes.}} have been computed by finding the position of the highest peaks in Fourier spectra of the corresponding Cartesian coordinates are $x$, $y$, $z$. 
\par
In addition to $f_x$, $f_y$, and $f_z$, we have determined the radial oscillations frequency $f_R$. Thus, a set of four numbers for every orbit in a disc for each model have been obtained. Using the ratios of these frequencies $f_x/f_\mathrm{R}$ and $f_y/f_\mathrm{R}$, \citet{Smirnov_etal2021} found several orbital groups in their grid of models. These orbital groups assemble in specific structures, and a dominant group or a combination of groups are responsible for a particular morphological feature of a model bar. Thus, the bar can be disassembled into the following orbital groups.
\par
\begin{itemize}
\item[-] A so-called `classic' bar, identified by the condition $f_R/f_x=2.0 \pm 0.1$. This orbital group primarily consists of loop orbits with $f_y/f_x=1 \pm 0.1$ (in the sense of the orbital classification by\citealt{Valluri_etal2016}) with x1-like and x2-like (prograde) or x4-like (retrograde) orbits being the prominent members of it. An x1-like family gives a narrow and elongated bar with an inner bar-like structure superimposed on an outer bar and oriented perpendicular to it (x2-like or x4-like orbits).
\item[-] Boxy orbits, selected via criterion $f_R/f_x=2.0 \pm 0.1$ and $f_y/f_x > 1.1$. These orbits constitute a face-on peanut, or the so called `boxy bar', and are part of the extended bar.
\item[-] Two non-classical groups: \blo and \blu in notation of \citet{Smirnov_etal2021}. 
\textcolor{black}{Orbits from these two groups have $f_R/f_x$ either larger than 2 (\blo) or smaller than 2 (\blu). They are definitely
not from the outer disc, but at the same time they are not quite elongated with
the bar and do not precess synchronously with it. These orbits have rather complex frequency proportions. Orbits from the \blu family have $f_y/f_x=1.0$. As to the orbits from the \blo family, they split into two branches, with $f_y/f_x=1.0$ (more extended structure) and with $(f_y+f_x)/f_R=1.0$ (more compact structure). 
All this gives them the overall `roundish' shape, which results in the formation of}
a barlens in the $N$-body model that has it. The \blu orbital group is a key component of a barlens and maintains the rounded shape in the central area of the bar. The \blo orbits are assembled into a more extended square-like structure, which is nevertheless a part of the barlens.
\end{itemize}
%%%%%%%%%%%%%%%%%%%%%%%%%%%%%%%%%%%%%%%%%%%%%%%%%%
%%%%%%%%%%%%%%%%%%%%%%%%%%%%%%%%%%%%%%%%%%%%%%%%%%
\par
The percentage of orbits of each type in the models is compiled in Table~\ref{tab:familiesnumbers}.
\par
One of the main conclusions made by \citet{Smirnov_etal2021} is that the dominance of one or another orbital group determines the face-on morphology of a bar and its features. In Section~\ref{sec:orbital_groups}, we use the data on the bar orbital groups revealed in \citet{Smirnov_etal2021} to determine their contribution to pixels with peanut signatures at the face-on position of the galaxy.

%%%%%%%%%%%%%%%%%%%%%%%%%%%%%%%%%%%%%%%%%%%%%%%%%%%%%%%%%%%%%%%%%%%%%%%
\begin{table}
\centering
\begin{tabular}{r|rrrrr}
\hline
{orbital} & {X} & {Xb} & {BLx} & {BL} & \\
{group} & & & & & \\
\hline
\hline
x1+x2+x4 & 3.79 & 6.52 & 9.51 & 11.35 & \\
boxy bar & 45.19 & 25.68 & 23.88 & 10.45 & \\
$\text{bl}_\text{o}$ & 5.36 & 7.82 & 9.98 & 13.49 & \\
$\text{bl}_\text{u}$ & 0.96 & 1.05 & 3.42 & 
 9.22 & \\
\hline
\end{tabular}
\caption{The percentage of orbits of each type in the models. The fraction (\%) is given relative to the total number of particles in the disc ($4\times 10^6$).}
\label{tab:familiesnumbers}
\end{table}
%%%%%%%%%%%%%%%%%%%%%%%%%%%%%%%%%%%%%%%%%%%%%%%%%%%%%%%%%%%%%%%%%%%%%%%

%%%%%%%%%%%%%%%%%%%%%%%%%%%%%%%%%%%%%%%%%%%%%%%%%%
%%%%%%%%%%%%%%%%%%%%%%%%%%%%%%%%%%%%%%%%%%%%%%%%%%
\subsection{Datacubes}
\label{sec:cubes}
% 2.3.
%%%%%%%%%%%%%%%%%%%%%%%%%%%%%%%%%%%%%%%%%%%%%%%%%%
%%%%%%%%%%%%%%%%%%%%%%%%%%%%%%%%%%%%%%%%%%%%%%%%%%
To study kinematics, one needs to represent the velocities of model particles in a way that can be faithfully compared with observational data. The most common form of data used in kinematic studies is a datacube. Here, we construct synthetic datacubes from our simulations.
\par
The best stellar kinematics available to date were obtained by TIMER project \citep{Gadotti_etal2019}, conducted on the MUSE spectrograph on the VLT. The high-level products of their processing pipeline are the maps with stellar kinematics parameters, which are publicly available on the TIMER website\footnote{\url{https://www.muse-timer.org}}. We use these maps as a reference point for the construction of our synthetic maps. 
We model our datacubes to have a comparable spatial resolution to MUSE. Specifically, we assume the whole image to be $30' \times 30'$ and the pixel size of each slice to be $300\text{px} \times 300\text{px}$. The proper choice of the spectral axis was more arbitrary. Normally, the velocities from integral-field spectroscopy (IFS) data are inferred via advanced pipelines, such as \texttt{GIST} \citep{Bittner_el_al2019}, which in turn uses \texttt{pPXF} \citep{Cappellari_Emsellem_2004} for stellar kinematics. These pipelines estimate the LOSVD by fitting a linear combination of template library spectra to the real spectrum of the galaxy. However, in case of our $N$-body models, a realistic imitation of stellar spectra would be unnecessary complex for the task at hand. Instead, we take the mean velocity uncertainty from \citep{Bittner_el_al2019}, where is was estimated to be $\approx 5 \text{km}\, \text{s}^{-1}$, and assume the same uncertainty for the positions of the LOSVD bins. Thus, we simply assume the bin width to be equal to  $2\sigma \approx 10 \text{km}/\text{s}$.  
\par
To improve signal-to-noise\footnote{In absence of real errors, we assume an artificial noise with Poisson distribution, thus S/N scales as $N^{-1/2}$.}
ratio (S/N) of the final datacubes, we stack 17 snapshots \textcolor{black}{in the time interval $t = 448$ -- $452$. During this period by $\Delta t = 4$, the change in the bar pattern speed is insignificant (see Figure~2 in \citealp{Smirnov_etal2021}).} \textcolor{black}{Thus},74.8 million particles in total \textcolor{black}{are reached}.
Contrary to the similar procedure in \citet{Iannuzzi_Athanassoula2015} we do not enforce any kind of symmetry to the model along the principal axes of the bar, because the total number of particles was already sufficient for the purposes of this paper after stacking.  
\par
To obtain the line-of-sight velocities, we rotate the model to match its $z$-axis, initially perpendicular to the disc plane, with the chosen direction of the line-of-sight and project the full velocity ($v$) on the new $z$-axis. Then, the velocity is binned from $\min v_{\text{LOS}}$ to $\max v_{\text{LOS}}$ with the bin width equal to the velocity resolution assumed above. In addition to that, we also compute the LOS projections of the vertical ($v_z$), azimuthal ($v_\varphi$), and radial ($v_R$) velocity components in the \emph{original} coordinate frame. Thus, for every spatial pixel of the datacube, we effectively have four distributions: one of the total LOS velocity and three distributions of the LOS velocity components. In practice, we just made a dedicated datacube for each velocity component to simplify the further processing. 
\par
In a similar fashion, for each model we prepare the set of datacubes excluding each orbital family one-by-one. In Section~\ref{sec:orbital_groups}, we  would use these datacubes to estimate the contributions of orbital groups in LOSVDs of datacubes. This approach was chosen instead of  considering each component alone, because it results in datacubes with a larger number of particles that are easier to analyse and to compare with observational data. 
\par
We deliberately do not apply the Voronoi-binning technique, which is quite common in similar studies, because in this case it harder to compare our datacubes for different models fiber-by-fiber (our Section~\ref{sec:LOSVD}) as well as different velocity components for the same model (Section~\ref{sec:LOSVD_vel_components}). In addition to that, it is rather unclear how to study the effect of individual orbital groups (our Section~\ref{sec:orbital_groups}), because the tessellation should be again different for every datacube and the additive property of LOSVD in a given fiber can no longer be used.  For the same reasons, we also do not smooth the layers of the datacubes with the instrumental PSF.
%%%%%%%%%%%%%%%%%%%%%%%%%%%%%%%%%%%%%%%%%%%%%%%%%%
%%%%%%%%%%%%%%%%%%%%%%%%%%%%%%%%%%%%%%%%%%%%%%%%%%
\subsection{Determination of kinematic parameters}
\label{sec:method}
% 2.4.
%%%%%%%%%%%%%%%%%%%%%%%%%%%%%%%%%%%%%%%%%%%%%%%%%%
%%%%%%%%%%%%%%%%%%%%%%%%%%%%%%%%%%%%%%%%%%%%%%%%%%
To describe deviations from the Gaussian profile, line-of-sight velocity distributions of galaxies are often represented by a product of its Gaussian and non-Gaussian parts. The latter term is further expanded into Gauss-Hermit series\citep{vanderMarel_Franx1993,Gerhard1993}, resulting in the following expression for the overall profile:
%%%%%%%%%%%%%%%%%%%%%%%%%%%%%%%%%%%%%%%%%%%%%%%%%%
%%%%%%%%%%%%%%%%%%%%%%%%%%%%%%%%%%%%%%%%%%%%%%%%%%
\begin{equation}
\mathcal{L}(v)= I \tfrac{1}{\sqrt{2\pi}} e^{-w^2 / 2} \cdot \sum\limits_{j=0}^{N}h_{j}H_{j}(v), \ w = \frac {v - \overline{V}}{\sigma},\\
\label{eq:fitting_generic}
\end{equation}
%%%%%%%%%%%%%%%%%%%%%%%%%%%%%%%%%%%%%%%%%%%%%%%%%%
%%%%%%%%%%%%%%%%%%%%%%%%%%%%%%%%%%%%%%%%%%%%%%%%%%
where $I$ is a normalization constant (corresponding to the intensity), $\overline{V}$ is a mean velocity, $\sigma$  is a velocity dispersion,  $H_{j}(v)$ are Hermite polynomials,  and $h_j$ are coefficient~(or factors) of Gauss-Hermit series.
\par 
The most commonly used parameters in kinematic studies  {apart from the mean velocity and its dispersion} are $h_3$ and $h_4$, which tell about the `skewness' and the `broadness' of the distribution respectively.
The values of these parameters can be estimated in different ways. One approach would be to use the orthogonality property of Hermit polynomials and compute the factors directly via the corresponding scalar products (see \citealp{Debattista_etal2005} for an example of its application). However, in this work, we follow \cite{vanderMarel_Franx1993,Brown_etal2013} and \cite{Iannuzzi_Athanassoula2015}, and simultaneously obtain the values for all free parameters ($I$, $\overline{V}$,  $\sigma$, $h_3$, and $h_4$). \textcolor{black}{This approach was chosen to ease the comparison with previous studies and for perfomance reasons.}
\textcolor{black}{Thus}, we truncate the series up to the fourth order to avoid overfitting to noise and effectively fit the profiles with the following function:
\begin{equation}
\mathcal{L}(v)=I \tfrac{1}{\sqrt{2\pi}} e^{-w^2 / 2}\Bigr( 1 + h_3 H_3(w) + h_4 H_4(w) \Bigl), \  w = \frac {v - \overline{V}}{\sigma}.  \\
\label{eq:fitting}
\end{equation}
In the case of a nearly Gaussian distribution, $h_3$ and $h_4$ parameters are around zero. However, 
if the system is dominated by radial orbits, the distribution becomes narrower compared to the Gaussian and the fit would produce positive values of $h_4$. If the system has predominantly circular orbits, then the distribution broadens. In this case, one measures the negative values of $h_4$ \citep{Gerhard1993,vanderMarel_Franx1993}. Negative values of $h_4$ may also appear when the LOSVD is a sum of two Gaussians with clear double-peaked structure, which is typical for two subsystems with different average rotation speeds \citep{Gerhard1993}.
\par
Using the fitting procedure described above\footnote{We additionally checked the fitting results with the MCMC parameter estimation and found no significant deviations from the former method, which is much faster.} (Eq.~\eqref{eq:fitting}), we obtain an intensity, mean velocity and its dispersion, and  $h_3$  and $h_4$ map for each synthetic datacube. 
To characterise  the density distribution, we calculate the $d_4$ parameter in a similar way using Eq.~(\ref{eq:fitting_d4}) 
%%%%%%%%%%%%%%%%%%%%%%%%%%%%%%%%%%%%%%%%%%%%%%%%%%
%%%%%%%%%%%%%%%%%%%%%%%%%%%%%%%%%%%%%%%%%%%%%%%%%%
\begin{equation}
\mathcal{L}(z)=\mu_0 \tfrac{1}{\sqrt{2\pi}} e^{-\psi ^2 / 2}\Bigr( 1 + d_3 H_3(\psi) + d_4 H_4(\psi) \Bigl),\  \psi=\frac{\mu_1-z}{\mu_2}\\
\label{eq:fitting_d4}
\end{equation}
%%%%%%%%%%%%%%%%%%%%%%%%%%%%%%%%%%%%%%%%%%%%%%%%%%
%%%%%%%%%%%%%%%%%%%%%%%%%%%%%%%%%%%%%%%%%%%%%%%%%%
with $(\mu_0, \mu_1, \mu_2, d_3, d_4)$ being the free parameters of the fit.

%%%%%%%%%%%%%%%%%%%%%%%%%%%%%%%%%%%%%%%%%%%%%%%%%%
%%%%%%%%%%%%%%%%%%%%%%%%%%%%%%%%%%%%%%%%%%%%%%%%%%
\section{LOSVD moment maps}
\label{sec:LOSVD}
% 3.
%%%%%%%%%%%%%%%%%%%%%%%%%%%%%%%%%%%%%%%%%%%%%%%%%%
%%%%%%%%%%%%%%%%%%%%%%%%%%%%%%%%%%%%%%%%%%%%%%%%%%
For galaxies with a B/PS bulge, the density distribution in the vertical direction in the area of the B/PS bulge differs from the usual exponential distribution or the distribution characteristic of a self-gravitating isothermal layer (see Eq.~\eqref{eq:rho_disk}). The density distribution is more vertically extended than the surrounding disc, i.e., the thickest parts of the bars are very loose, and the parameter $d_4$ in Eq.~\eqref{eq:fitting_d4}) becomes negative. 
In this section, we follow the approach of \citet{Debattista_etal2005} and explore the connection between vertical density features~($d_4$) and the thickness with the kinematics~($h_4$) for different inclinations of the model and different bar position angles.

%%%%%%%%%%%%%%%%%%%%%%%%%%%%%%%%%%%%%%%%%%%%%%%%%%
%%%%%%%%%%%%%%%%%%%%%%%%%%%%%%%%%%%%%%%%%%%%%%%%%%
\subsection{Face-on maps}
\label{sec:LOSVD_i0}
% 3.1.
%%%%%%%%%%%%%%%%%%%%%%%%%%%%%%%%%%%%%%%%%%%%%%%%%%
%%%%%%%%%%%%%%%%%%%%%%%%%%%%%%%%%%%%%%%%%%%%%%%%%%

%%%%%%%%%%%%%%%%%%%%%%%%%%%%%%%%%%%%%%%%%%%%%%%%%%
%%%%%%%%%%%%%%%%%%%%%%%%%%%%%%%%%%%%%%%%%%%%%%%%%%
% Fig.~2.
\begin{figure*}
\centering
\includegraphics[width=1\linewidth]{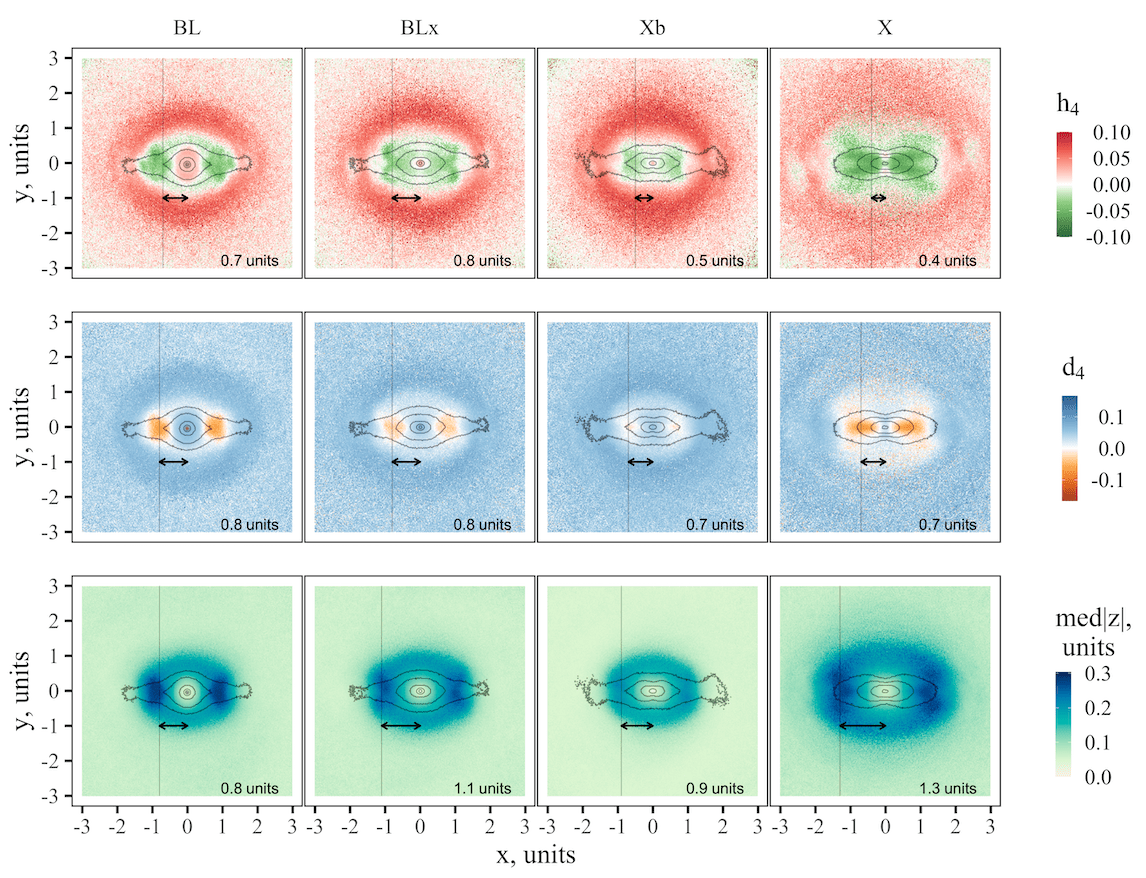}
\caption{2D maps of the $h_4$ parameter~(\textit{upper} row), the vertical density distribution parameter $d_4$~(\textit{middle} row) and the $\med|z|$~(\textit{bottom} row) for all model galaxies at face-on position ($i=0^\circ$, $\mathrm{PA}=0^\circ$). The isophotes for each model correspond to the intensity maps shown in Fig.~\ref{fig:models_view}. The vertical lines define the position of the $h_4$ and $d_4$ minima and the maximum of thickness (med|z|) along the major axis of the bar~(the exact values of the distance from the centre of the model galaxy to the extremum are indicated at the bottom of the maps).}%
\label{fig:2Dmaps_i0PA0}
\end{figure*}
%%%%%%%%%%%%%%%%%%%%%%%%%%%%%%%%%%%%%%%%%%%%%%%%%
%%%%%%%%%%%%%%%%%%%%%%%%%%%%%%%%%%%%%%%%%%%%%%%%%%

Fig.~\ref{fig:2Dmaps_i0PA0}~(upper row) demonstrates 2D maps of the parameter $h_4$ for our models (BL, BLx, Xb, and X). We obtained $h_4$ values for each pixel constituting the face-on snapshots of models as described in Section~\ref{sec:method}. The $h_4$ maps for prepared models have different morphology for the face-on case due to the different morphology of the bar.	
\par
Regardless of the type of the B/PS bulge, for all modelled maps there are negative values of $h_4$ within the bar radii~(Fig.~\ref{fig:2Dmaps_i0PA0}). The cuts along the major axis also display the double-sided minima as discussed in \cite{Debattista_etal2005}. Although, we note that the difference between these $h_4$ maps depends on the exact morphology of the model. 
%The BL model $h_4$ map differs from others.   Also, t
For the BL model, the minima of the $h_4$ parameter lie outside the rounded central isophotes, and the pixels with negative values of $h_4$ are grouped mainly on the major axis of the bar.
At the same time, for BLx, Xb, and X models, the negative values of $h_4$ parameter draw an area that looks like a `bow tie'.
In the particular case of the model X, which has the largest B/PS bulge of all of the models,  the negative values of the parameter $h_4$ still lie close to the site occupied by the face-on peanut-shaped bar, while the morphology of the $h_4$ maps reflects the X-shaped morphology of the its bar (compare Fig.~\ref{fig:models_view} and Fig.~\ref{fig:2Dmaps_i0PA0}). 
\par
\begin{figure}
\centering
\includegraphics[width=1\linewidth]{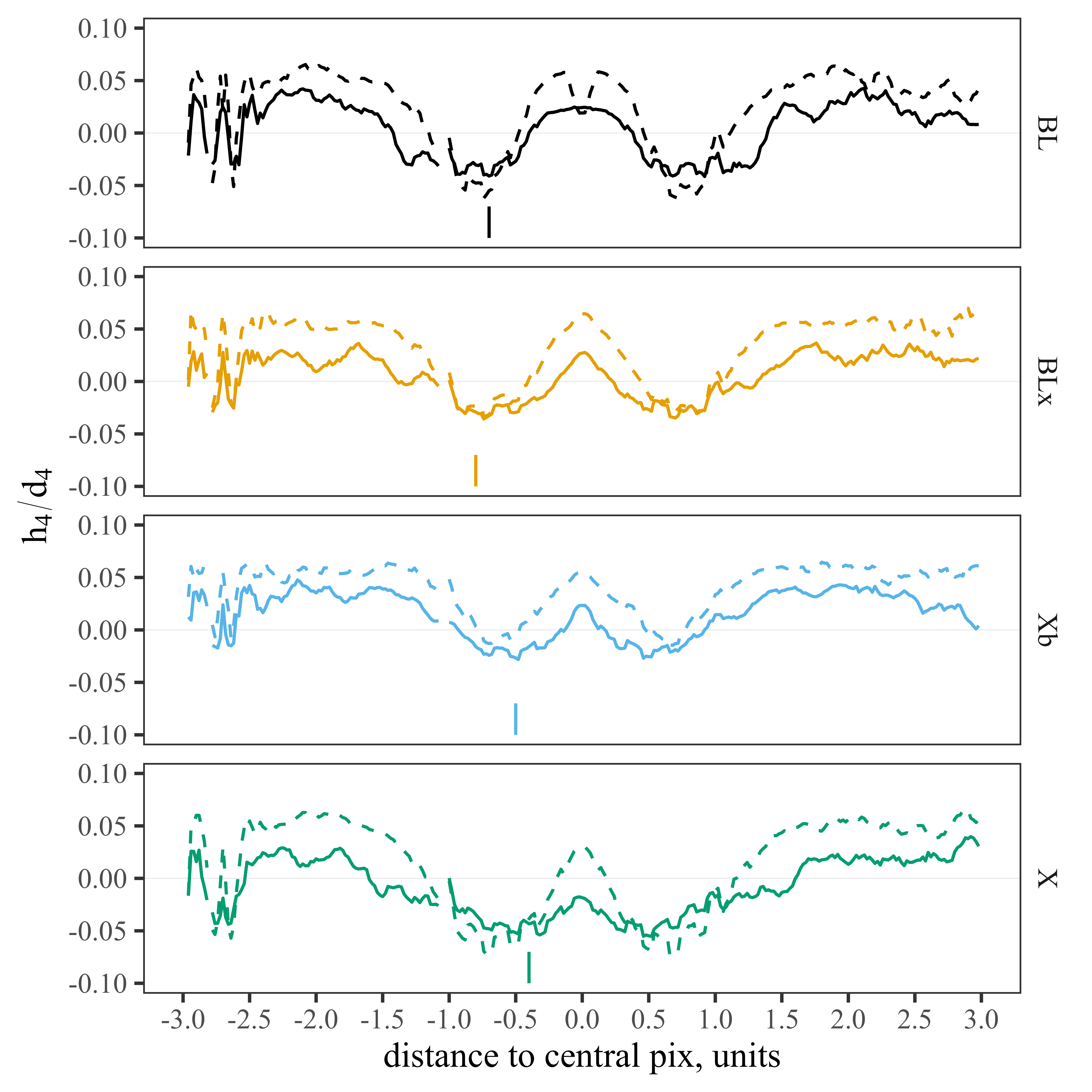}
\caption{The $h_4$~(solid lines) and $d_4$~(dashed lines)
%and med|$z$| (dotted line) 
profiles along the major axis for the BL, BLx, Xb, and X models. The short vertical bars at the bottom of the plots indicate the position of the $h_4$ minima and correspond to the vertical lines in Fig.~\ref{fig:2Dmaps_i0PA0} (upper row).}%
\label{fig:h4_alongmajor_inc0pa0}
\end{figure}

Using cuts along the major axis of the bar (Fig.~\ref{fig:h4_alongmajor_inc0pa0}, solid lines), we show the values of the $h_4$ parameter along the major axis for all models. In Fig.~\ref{fig:2Dmaps_i0PA0} (upper row) the position of $h_4$ minimum is shown by a vertical line and its location is indicated in the dimensionless units of this work. In all models, the minimum values of $h_4$  are similar and equal to $-0.05$. However, the minimum of $h_4$ along the major axis of the bar is reached at different distances from the centre. For the BL and BLx models, the $h_4$ minimum along the major axis coincides with the boundary of inner rounded isophotes where the transition to a narrow extended bar occurs. Xb and X models have their $h_4$ minima along the major axis closer to the centre of the galaxy than the BL and BLx models, but they are wide~(especially for the X model). This is clearly seen on the profiles of the $h_4$ parameter along the major axis of the bar (Fig. \ref{fig:h4_alongmajor_inc0pa0}). If we compare the position of the negative minima of the parameter $h_4$ on the major axis with the dimensions of the B/PS bulge, as it is seen on the side-on snapshots (Fig.~\ref{fig:models_view}), then they do not match. B/PS bulge is larger. At the same time, the maps in the Fig. \ref{fig:2Dmaps_i0PA0} (upper row) show that negative values of the $h_4$ parameter are present not only along the major axis of the bar, but are also partially localized in the bar itself away the minima on the major axis along four rays. The rays roughly outline the dimensions of the B/PS bulge.
\par
We also checked the relationship between the flat-topped distribution of the LOSV (measured by $h_4$) and the features of the vertical density distribution~($d_4$), as well as between $h_4$ and the thickness for all our models. To measure the thickness, we calculate the median value of absolute distance to the disc plane (hereinafter, $\med |z|$) for all particles that create a signal in a given pixel. We also computed $d_4$ using Eq.~\eqref{eq:fitting_d4}. The middle and bottom rows in Fig.~\ref{fig:2Dmaps_i0PA0} show the $d_4$ and $\med|z|$, respectively, for all our models. Regardless of a specific model, all our models have a good correlation between $h_4$ and $d_4$ parameters, as galaxies with B/PS bulges. We additionally checked the Pearson correlation coefficient between $h_4$ and $d_4$ along the major axis. For our models, it varies from 0.79 to 0.83.
%\par
\par
The Xb model is a little out of the general range of models. It has the loosest and most massive classical bulge. In the area of the bar and at the face-on position, the bulge particles through which the line of sight passes blur the features of the vertical density distribution of the B/PS bulge, which, for this reason, hardly appear on the $d_4$ map (Fig.~\ref{fig:2Dmaps_i0PA0}, middle row). At the same time, B/PS bulge signatures are still visible on the kinematic maps ($h_4$ maps, Fig.~\ref{fig:2Dmaps_i0PA0}, upper row), and green pixels (negative values) are located along the bar up to the boundaries of the B/PS bulge, as seen edge-on (Fig.~\ref{fig:models_view}). 
\par
Thus, the flat-topped vertical density distribution can be traced from the flat-topped LOSVD distribution for face-on galaxies. In the following sections, we discuss the limitations of this approach.
\par
The relationship between $h_4$ and the thickness ($\med|z|$) of the B/PS bulge is shown in Fig.~\ref{fig:2Dmaps_i0PA0} (upper and bottom rows). We have obtained a formal match between the minima of the $h_4$ parameter and the thickness only for the BL model. For all other B/PS bulge models, the $h_4$ parameter does not trace the thickness. Thus, the parameter $h_4$ is only an indicator of features of the vertical density distribution, but not the thickness.

%%%%%%%%%%%%%%%%%%%%%%%%%%%%%%%%%%%%%%%%%%%%%%%%%%
%%%%%%%%%%%%%%%%%%%%%%%%%%%%%%%%%%%%%%%%%%%%%%%%%%
% Fig.~3.
%%%%%%%%%%%%%%%%%%%%%%%%%%%%%%%%%%%%%%%%%%%%%%%%%%
%%%%%%%%%%%%%%%%%%%%%%%%%%%%%%%%%%%%%%%%%%%%%%%%%%

%%%%%%%%%%%%%%%%%%%%%%%%%%%%%%%%%%%%%%%%%%%%%%%%%%
%%%%%%%%%%%%%%%%%%%%%%%%%%%%%%%%%%%%%%%%%%%%%%%%%%
\subsection{Inclination effects for $h_4$ maps}
\label{sec:LOSVD_inclination_bl}
% 3.2.
%%%%%%%%%%%%%%%%%%%%%%%%%%%%%%%%%%%%%%%%%%%%%%%%%%
%%%%%%%%%%%%%%%%%%%%%%%%%%%%%%%%%%%%%%%%%%%%%%%%%%
In this subsection, we explore in detail how  
the $h_4$ maps change depending on the galaxy inclination and the position angle (PA) of the bar major axis relative to the line of nodes. We also discuss the absence of connection between $h_4$ and $d_4$ when the model is inclined. 
We now focus only on the BL model, since all other models behave in a similar way. The corresponding figures for BLx, Xb, and X model can be found in Appendix~\ref{app:inclination_full_maps}. 
In Fig.~\ref{fig:BL_all_i} we present the changes of $h_4$ maps varying both inclination angles ($0^\circ$, $20^\circ$, $40^\circ$, $60^\circ$) and position angles~($0^\circ$, $45^\circ$, $90^\circ$) of the bar. This set of observer positions broadly covers all relevant options. We additionally add inset plots with $d_4$ maps to the upper right corner of each panel with $h_4$ maps. 
\par
The leftmost columns of Fig.~\ref{fig:BL_all_i} show $h_4$~(main panels) and $d_4$~(inset panels) parameters for different inclinations with $\mathrm{PA} = 0^{\circ}$. The parameter $d_4$ stops being a reliable vertical density distribution indicator with the inclination of a model galaxy even if the position angle is close to zero. It happens because particles at different distances from the center of the galaxy end up in the same line of sight. Respectively, the connection between $h_4$ and $d_4$ is broken. The reason is the contamination of $v_z$ by other velocity components~(we discuss it in detail in Sec.~\ref{sec:LOSVD_vel_components}).
This is especially well seen on the $h_4$ and $d_4$ profiles along the major axis of a bar (Fig.~\ref{fig:h4_alongmajor_all_incpa0}), where the  $h_4$ parameter along the major axis behaves in a different way than the $d_4$. The inclination of the galaxy moves the minima of $h_4$ away from the galaxy center. Also, for inclined galaxies~($ i \gtrsim 20^{\circ}$), the $h_4$ minima become less prominent. We support this statement by the cuts along the major axis for the BL model in Fig.~\ref{fig:h4_alongmajor_all_incpa0}. 
\par
The same tendency to move the minimum of $h_4$ to the periphery along the major axis with increasing inclination is also observed when the inclined galaxy has a non-zero position angle. The second and the third columns of Fig.~\ref{fig:BL_all_i} show the $h_4$ for $\mathrm{PA}=45^\circ$ and $\mathrm{PA}=90^\circ$ respectively. 
The prominent two-sided $h_4$ minima along the major axis create a semi-closed `ring' of negative values already at $\mathrm{PA}=45^\circ$ and closed `ring' at $\mathrm{PA}=90^\circ$ . Maps with $i\approx$ 40$^{\circ}$-60$^{\circ}$ look very disturbed and the deepness of $h_4$ minima on the line of nodes increases~(Fig.~\ref{fig:h4_alongmajor_all_incpa0}).
%A stronger rotation of 
A change of the PA for an inclined galaxy also increases the $h_4$ minima, and creates a prominent `ring'\footnote{Signatures of a ring from negative values of $h_4$ can be noticed in figure~29 by \citet{Iannuzzi_Athanassoula2015} (the case $\mathrm{PA}=90^\circ$, $i=60^\circ$, models GTR101 and GTR116, which have barlenses).} of $h_4$ minima. Moreover, such a ring is typical not only for the model with barlens~(BL) but also for all the others~(Fig.~\ref{fig:BLx_all_i}, ~\ref{fig:Xb_all_i}, ~\ref{fig:X_all_i}).
\par
When $\mathrm{PA}=0^\circ$ and $i\geq40^\circ$, the minima of $h_4$ lie on the major axis of the bar, and the minima of $d_4$ form two arcs symmetrically crossing the minor axis. At the same time, for $\mathrm{PA}=90^\circ$ and $i\geq40^\circ$, the deepest minima of the parameter $h_4$ appear along the minor axis, which are not reflected in any way in the minima of the parameter $d_4$, although the relationship between the minima can be traced along the major axis. We return to the discussion of these features in Sections~\ref{sec:LOSVD_vel_components} and~\ref{sec:orbital_groups}. 
\par
To conclude this section, we note that the appearance of $h_4$ maps is affected not only by the actual morphology of the bar but also by the position of the barred galaxy  with respect to the observer. We note that for all our models with B/PS bulges, these effects manifest themselves in a similar way. At the same time, for inclined galaxies~( $i\geq20^\circ$), the features of the LOSVD are no longer related to the features of the density distribution in the vertical direction. In the next subsection, we compare all our models at an intermediate inclination in an attempt to better characterise the features of the maps. We also pay attention to the fact that when the galaxy is inclined, the minima move away from the center of the model, and their position can no longer reflect the size of the bar.

%%%%%%%%%%%%%%%%%%%%%%%%%%%%%%%%%%%%%%%%%%%%%%%%%%
%%%%%%%%%%%%%%%%%%%%%%%%%%%%%%%%%%%%%%%%%%%%%%%%%%
% Fig.~4.
\begin{figure}
\centering
\includegraphics[width=1.0\linewidth]{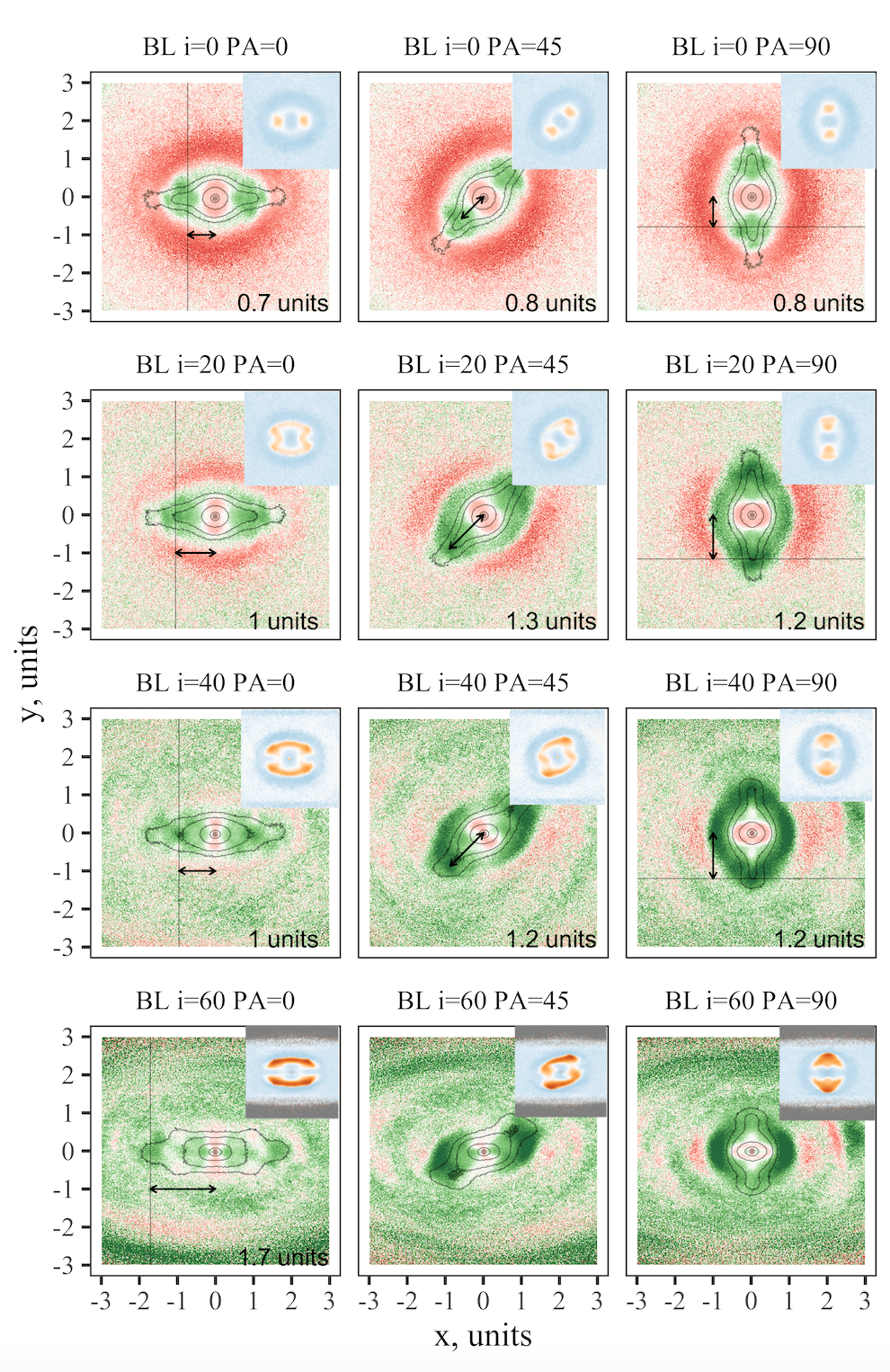}
\caption{Effects of galaxy inclination ($i$) and PA in the $h_4$ maps for the barlens model~(BL). The isophotes correspond to the intensity maps for a given inclination. Straight lines with arrows show the position of the $h_4$ minima along the major axis of the bar. The number at the bottom shows the distance from the centre of the model galaxy to the $h_4$ minimum along the major axis. For $i=60^\circ$ and $\mathrm{PA}=45^\circ, \, 90^\circ$ the positions of the minima are not indicated since there are no global minima along the major axis. In the upper right corner of each panel, the  insets with $d_4$ corresponding to the given position of the galaxy are added. The colors are the same as in Fig.~\ref{fig:2Dmaps_i0PA0}. \textcolor{black}{The minima of $h_4$ are broad. Therefore, the precise location of exact global minima is not very well defined and may slightly differ for different viewing angles even in case of $i=0^\circ$.}}
\label{fig:BL_all_i}
\end{figure}
%%%%%%%%%%%%%%%%%%%%%%%%%%%%%%%%%%%%%%%%%%%%%%%%%%
%%%%%%%%%%%%%%%%%%%%%%%%%%%%%%%%%%%%%%%%%%%%%%%%%%

%%%%%%%%%%%%%%%%%%%%%%%%%%%%%%%%%%%%%%%%%%%%%%%%%%
%%%%%%%%%%%%%%%%%%%%%%%%%%%%%%%%%%%%%%%%%%%%%%%%%%
% Fig.~5.
\begin{figure*}
\centering
\includegraphics[width=1\linewidth]{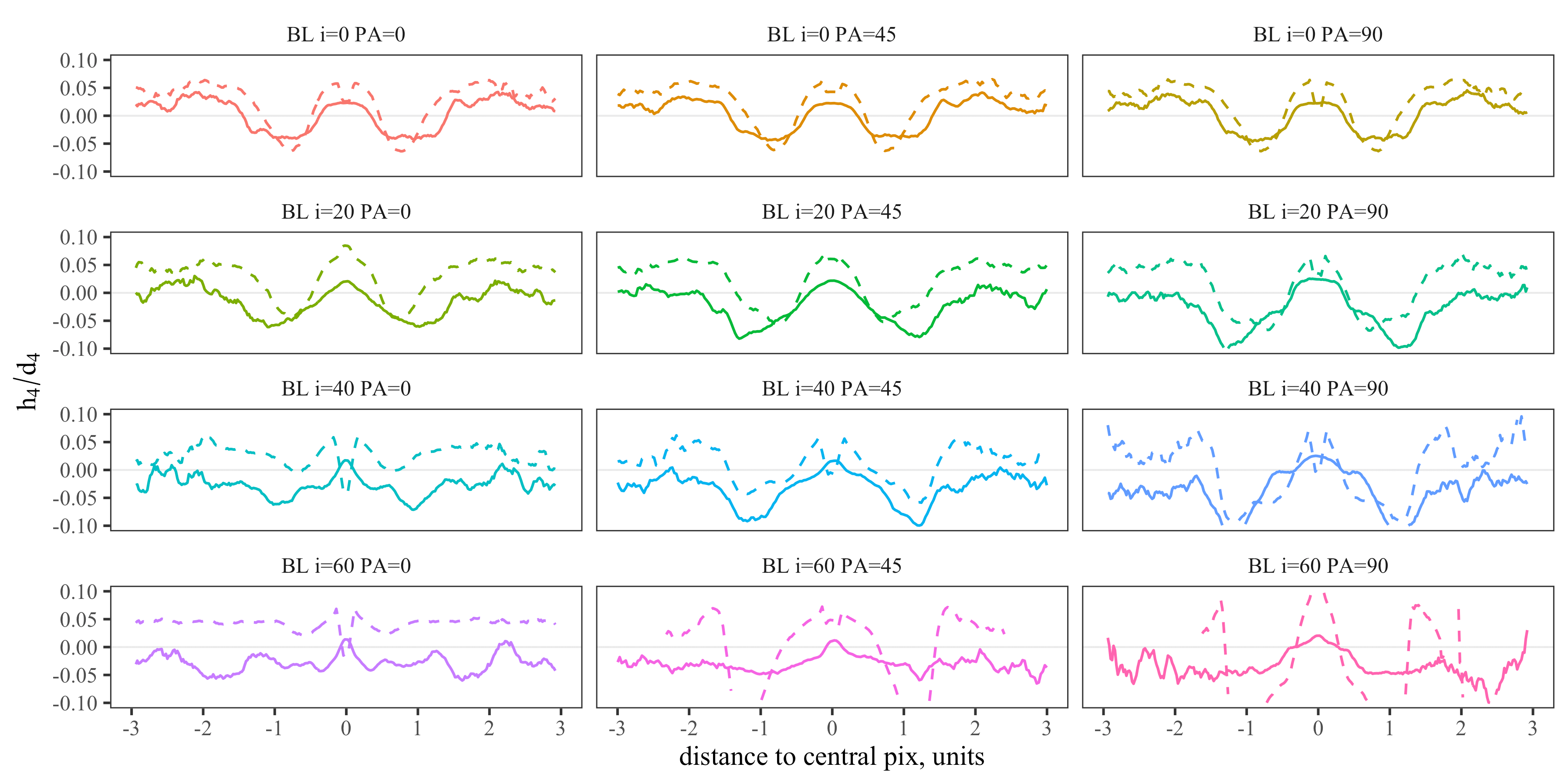}
\caption{The $h_4$~(solid lines) and $d_4$~(dashed lines) profiles along the major axis for the BL model at different values of $i$ and PA.}%
\label{fig:h4_alongmajor_all_incpa0}
\end{figure*}
%%%%%%%%%%%%%%%%%%%%%%%%%%%%%%%%%%%%%%%%%%%%%%%%%%
%%%%%%%%%%%%%%%%%%%%%%%%%%%%%%%%%%%%%%%%%%%%%%%%%%

%%%%%%%%%%%%%%%%%%%%%%%%%%%%%%%%%%%%%%%%%%%%%%%%%%
%%%%%%%%%%%%%%%%%%%%%%%%%%%%%%%%%%%%%%%%%%%%%%%%%%
\subsection{The differences between models on $h_4$ maps}
\label{sec:LOSVD_inclination_between_models}
\begin{figure}
\centering
\includegraphics[width=1.0\linewidth]{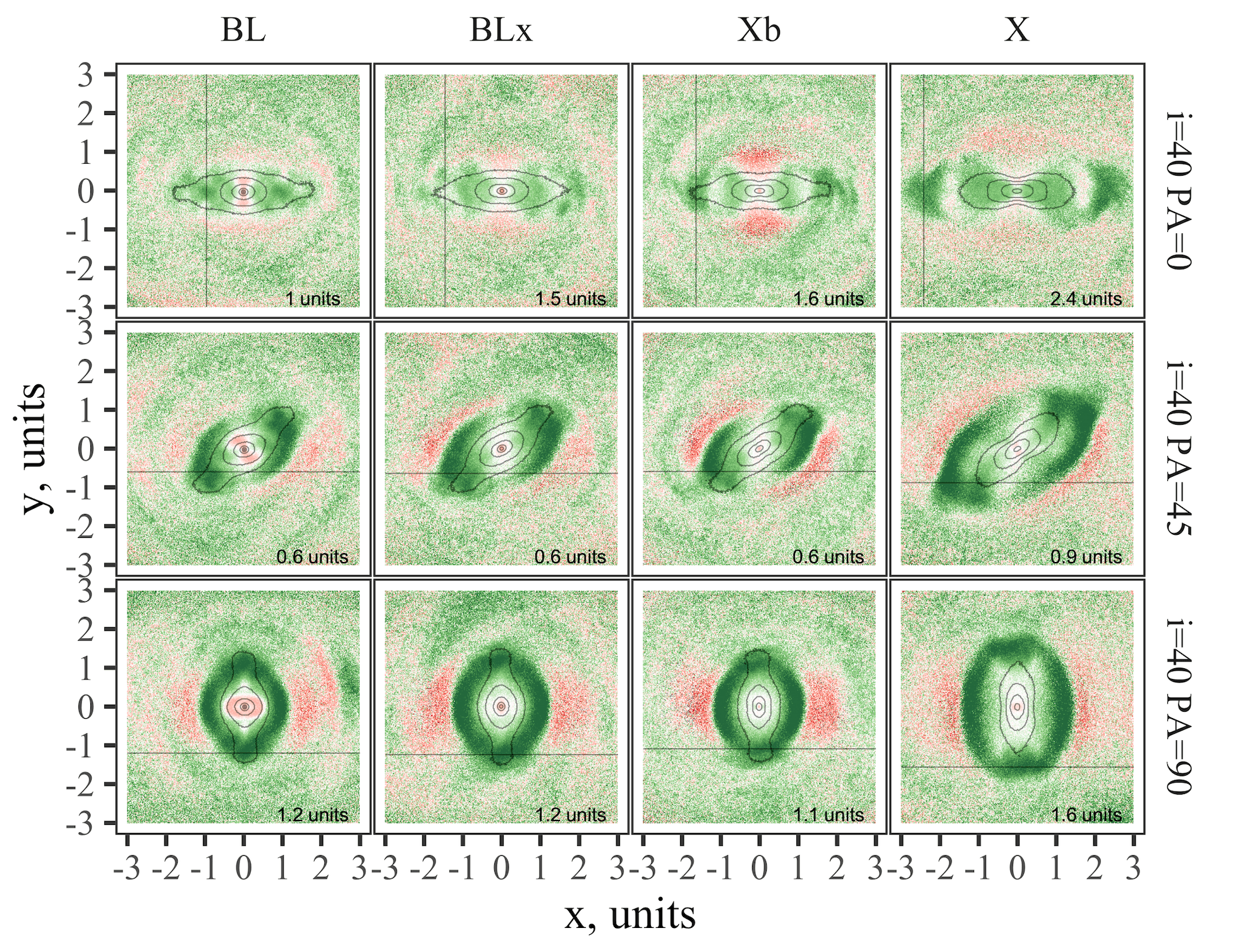}
\caption{Differences between $h_4$ maps for all models~(BL, BLx, Xb, X) at intermediate inclination $i=40^{\circ}$ and varying position angles. The isophotes correspond to the intensity maps for a given position of the model relative to the observer. The vertical lines in the top row indicate the positions of $h_4$ minima along the major axis. The exact numbers at the bottom right corner of each panel show the distance from the centre of the model galaxy to the $h_4$ minima. The colors are the same as in Fig.~\ref{fig:2Dmaps_i0PA0}.}
\label{fig:incpa_affect_example_models}
\end{figure}

% 3.3.
%%%%%%%%%%%%%%%%%%%%%%%%%%%%%%%%%%%%%%%%%%%%%%%%%%
%%%%%%%%%%%%%%%%%%%%%%%%%%%%%%%%%%%%%%%%%%%%%%%%%%
Many galaxies from the TIMER project \citep{Gadotti_etal2019} have inclinations in the range $i=30^\circ-60^\circ$. 
So, in this section, we pay special attention to the differences between our models with a significant inclination.  Fig.~\ref{fig:incpa_affect_example_models} shows the $h_4$ maps for all models at  $i=40^{\circ}$ and different position angles $\mathrm{PA}=0^\circ\,, 45^\circ\,, 90^{\circ}$. The first line corresponds to $\mathrm{PA}=0^\circ$ and we do not notice a significant difference between the models BL, BLx, and Xb. Although the map for the model without the bulge (X) differs from the others and the position of its $h_4$ minimum along the major axis of the bar lies as far as possible from the centre of the galaxy~(2.4 units), 
%we do not believe that 
it does not make it possible to distinguish between the vertical structure of bars of different morphology, i.e., to identify B/PS bulges with certain parts of the bar.
\par
The rotation of the kinematic axis by $45^\circ$~(second row in Fig.~\ref{fig:incpa_affect_example_models}) makes the BL, BLx and Xb models even less visually distinguishable. The main difference between the $h_4$ maps for these models at such position angles is the degree of compactness of the location of the features on the maps: the minimum of $h_4$ is slightly further from the centre for the BLx model. The differences between the BL and Xb models are completely insignificant, although the Xb model does not contain a barlens. As for the X model, we note that the $h_4$ maps corresponding to this model differ from the rest only in that the minima are located almost one and a half times farther from the centre of the galaxy~(0.9 units versus 0.6), due to the fact that the bar of this model is more extended, as well as its B/PS bulge (cf. Fig.~\ref{fig:models_view}).
\par
In the case $\mathrm{PA}=90^\circ$, the map shows the ring of negative values of $h_4$ for all models, regardless of the existence or absence of a barlens. The main difference between the rings of negative values of $h_4$ for the BL and X models is their shape~(for the X model the ring is more squared) and the size of the ring~(for the X model the ring is larger). However, there is practically no difference between the BL, BLx, and Xb models for this position of the galaxy.
\par
Thus, the presence of the ring of negative values of $h_4$ is not related to the exact bar morphology, but to the inclination effects, which can manifest themselves in two ways. First, at intermediate inclinations, the LOSVD is contaminated by other velocity components, in addition to $v_z$. Thus,  the negative values of $h_4$ are determined by the contribution of these other components to the LOSVD and their features, and not by the looseness of the density distribution. 
Second, stars with other kinematics, different from the kinematics of the stars of the bar or any of its subsystems, may appear on the line of sight. Even if the VD for individual subsystems is close to a Gaussian, the sum of Gaussians with different mean velocities can lead to the effect of negative values of $h_4$ \citep{Gerhard1993}.

%%%%%%%%%%%%%%%%%%%%%%%%%%%%%%%%%%%%%%%%%%%%%%%%%%
%%%%%%%%%%%%%%%%%%%%%%%%%%%%%%%%%%%%%%%%%%%%%%%%%%
% Fig.~6.
%%%%%%%%%%%%%%%%%%%%%%%%%%%%%%%%%%%%%%%%%%%%%%%%%%
%%%%%%%%%%%%%%%%%%%%%%%%%%%%%%%%%%%%%%%%%%%%%%%%%%

%%%%%%%%%%%%%%%%%%%%%%%%%%%%%%%%%%%%%%%%%%%%%%%%%%
%%%%%%%%%%%%%%%%%%%%%%%%%%%%%%%%%%%%%%%%%%%%%%%%%%
\section{Contamination of the LOSVD by different velocity components}
\label{sec:LOSVD_vel_components}
% 4.
%%%%%%%%%%%%%%%%%%%%%%%%%%%%%%%%%%%%%%%%%%%%%%%%%%
%%%%%%%%%%%%%%%%%%%%%%%%%%%%%%%%%%%%%%%%%%%%%%%%%%
In Section~\ref{sec:LOSVD}, we only discussed how the $h_4$ maps generally change for our four models depending on the bar morphology, inclination, and position angle of a bar major axis relative to the line of nodes. In this section, our goal is to explain why the location of negative $h_4$ minima for our BL model shifts on the maps depending on the inclination and position angle. To do this, we consider only the BL model and separate the velocity components in each cube since now. In contrast to the face-on position, for inclined galaxy, the LOSVDs can contain contributions from both $v_\varphi$ and $v_{\mathrm{R}}$ in addition to $v_z$. These velocity components distort the LOSVD, although they have nothing to do with the flat-topped density distribution in the vertical direction.

%%%%%%%%%%%%%%%%%%%%%%%%%%%%%%%%%%%%%%%%%%%%%%%%%%
%%%%%%%%%%%%%%%%%%%%%%%%%%%%%%%%%%%%%%%%%%%%%%%%%%
\subsection{Contamination of LOSVD by azimuthal velocity component}
\label{sec:LOSVD_vphi}
% 4.1.
%%%%%%%%%%%%%%%%%%%%%%%%%%%%%%%%%%%%%%%%%%%%%%%%%%
%%%%%%%%%%%%%%%%%%%%%%%%%%%%%%%%%%%%%%%%%%%%%%%%%%

%%%%%%%%%%%%%%%%%%%%%%%%%%%%%%%%%%%%%%%%%%%%%%%%%%
%%%%%%%%%%%%%%%%%%%%%%%%%%%%%%%%%%%%%%%%%%%%%%%%%%
% Fig.~7.
\begin{figure}
\centering
\includegraphics[width=1.0\linewidth]{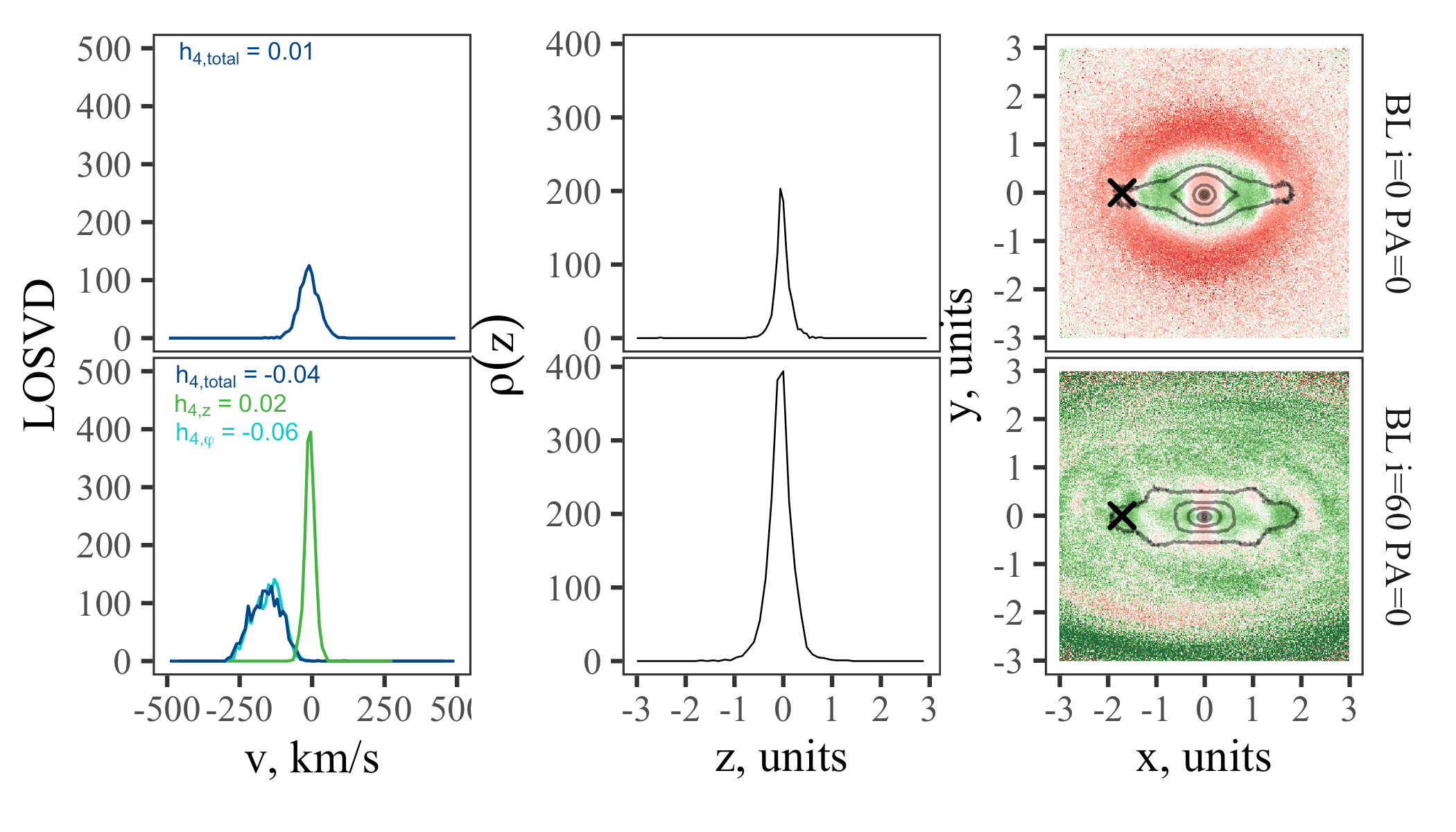}
\caption{\textit{Left:} LOSVDs for the pixel (-1.7; 0) of the BL model at face-on position~(top) and for the same model but inclined at $i=60^{\circ}$~(bottom) with respect to the bar major axis.  The total LOSV (dark blue color) was additionally decomposed  into vertical~($v_z$ , green color) and azimuthal ($v_\varphi$, cyan) components.
The selected pixel corresponds to the $h_4$ minima along the major axis for the BL model at $i=60^{\circ}$. \textit{Centre:} the vertical density distribution for both inclinations. \textit{Right:} $h_4$ maps for $i=0^{\circ}$~(top) and $i=60^{\circ}$~(bottom). The isophotes~(black lines) correspond to the intensity maps for a given inclination. The cross symbols show the position of selected pixels.}
\label{fig:h4_velcontrib_pa0}
\end{figure}
%%%%%%%%%%%%%%%%%%%%%%%%%%%%%%%%%%%%%%%%%%%%%%%%%%
%%%%%%%%%%%%%%%%%%%%%%%%%%%%%%%%%%%%%%%%%%%%%%%%%%

%%%%%%%%%%%%%%%%%%%%%%%%%%%%%%%%%%%%%%%%%%%%%%%%%%
%%%%%%%%%%%%%%%%%%%%%%%%%%%%%%%%%%%%%%%%%%%%%%%%%%
% Fig.~8.
\begin{figure}
\centering
\includegraphics[width=\linewidth]{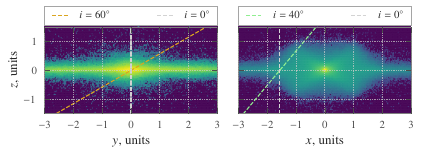}
\caption{\textit{Left}: a $(yz)$ projection of the region selected by $|x+1.7|<0.02$ (i.e., a thin layer); LOS rays have angles $i=0^\circ$ (light gray) and $i=60^\circ$ (orange) and pass through the zero point of this view. \textit{Right}: an $(xz)$ projection of the layer $|y|< 0.02$; LOS rays have angles $i=0^\circ$ (light gray) and $i=40^\circ$ (light green) and pass through the point $(x,z) = (-1.57, 0)$.}%
\label{fig:BL_layers_los}
\end{figure}
%%%%%%%%%%%%%%%%%%%%%%%%%%%%%%%%%%%%%%%%%%%%%%%%%%
%%%%%%%%%%%%%%%%%%%%%%%%%%%%%%%%%%%%%%%%%%%%%%%%%%

%%%%%%%%%%%%%%%%%%%%%%%%%%%%%%%%%%%%%%%%%%%%%%%%%%
%%%%%%%%%%%%%%%%%%%%%%%%%%%%%%%%%%%%%%%%%%%%%%%%%%
Our first example is a pixel on the bar major axis for the BL model, at the distance of 1.7 units from the centre (Fig.~\ref{fig:h4_velcontrib_pa0}, upper right plot, a cross). In this pixel, particles with different $z$, but with the nearly same values of $y=0$ (for $i=0^\circ$), fall on the line of sight (Fig.~\ref{fig:BL_layers_los}, left plot, light grey line). In the face-on position, when only the vertical component of the velocity $v_z$ contributes to the LOSVD, the value of $h_4$ in this pixel is positive (Fig.~\ref{fig:h4_alongmajor_all_incpa0}, upper left cut, solid line and Fig.~\ref{fig:h4_velcontrib_pa0}, upper right plot, a cross in red region), and the distribution function $f(v_z)$, in contrast to the Gaussian, has a sharp peak (Fig.~\ref{fig:h4_velcontrib_pa0}, upper left plot). This is in good agreement with the density distribution in the vertical direction, which also has a narrow peak (Fig.~\ref{fig:h4_velcontrib_pa0}, upper middle plot).
\par
When the stellar disc is inclined around the major axis of the bar at an angle $i=60^\circ$, the distance from this pixel to the centre of the disc does not change (Fig.~\ref{fig:h4_velcontrib_pa0}, bottom right plot, a cross). At the same time, the line of sight in this case captures particles not only with different $z$ but also with different $y$ (Fig.~\ref{fig:BL_layers_los}, left plot, orange line). Now, not only the vertical velocity component $v_z$ (green line on the bottom left panel of Fig.~\ref{fig:h4_velcontrib_pa0}), but also the azimuthal one (cyan line), contributes to the LOSVD in this pixel, and the LOSVD (dark blue line) is characterised by a negative value of $h_4$ (Fig.~\ref{fig:h4_velcontrib_pa0}, bottom right plot, a cross in green area). Moreover, in this pixel, there is a negative minimum of $h_4$, which was not in this place for the model in the face-on position (Fig.~\ref{fig:h4_alongmajor_all_incpa0}, left cuts, solid lines). The LOSVD in this pixel is flat-topped (Fig.~\ref{fig:h4_velcontrib_pa0}, bottom left plot) but its shape is determined mainly by $f(v_\varphi)$ because $f(v_z)$ in this pixel is narrow peak around zero, which means that the projection of the $v_z$ on the LOS is rather small . We know that it is the shape of the distribution function $f(v_z)$ that is related to the features of the density distribution in the vertical direction, i.e., only a flat-topped $f(v_z)$, which gives a negative value of $h_4$, implies a flat-topped density distribution. Thus, it is impossible to conclude that this pixel shows the signatures of a peanut. This is confirmed by the density distribution along the line of sight (Fig.~\ref{fig:h4_velcontrib_pa0}, bottom middle plot), although formally the value of $h_4$ in this pixel falls into negative minimum.
\par
At smaller inclinations ($i=20^\circ-40^\circ$), pixels with negative values of $h_4$ along the bar major axis are located approximately in the same area as for the face-on view (Fig~\ref{fig:BL_all_i}, left column), especially considering that the negative minimum of $h_4$ for the face-on view is quite wide. This is clearly seen in the cuts along the major axis for different inclinations (Fig.~\ref{fig:h4_alongmajor_all_incpa0}, left column, solid lines). As mentioned above, the position of the formal negative minimum of $h_4$ moves outward. At the same time, at $i=20^\circ$, in the region of the negative minimum of $h_4$, there are still pixels with negative values of $d_4$ along the major axis, and there is a pronounced negative minimum of $d_4$, although it does not coincide with the position of the $h_4$ minimum. At $i=40^\circ$, pixels with negative values of $d_4$ disappear completely, and the negative minimum of $h_4$ is strongly blurred in the B/PS bulge region. This means that for the low inclinations ($i<40^\circ$) and $\mathrm{PA}=0$, negative values of $h_4$ along the major axis carry the information about the presence of a peanut despite the contamination of the LOSVD with the $v_\varphi$ velocity component. However, this is no longer the case for larger inclinations of the galaxy even on the major axis of the bar. 

%%%%%%%%%%%%%%%%%%%%%%%%%%%%%%%%%%%%%%%%%%%%%%%%%%
%%%%%%%%%%%%%%%%%%%%%%%%%%%%%%%%%%%%%%%%%%%%%%%%%%
\subsection{Contamination of LOSVD by radial velocity component}
\label{sec:LOSVD_vR}
% 4.2.
%%%%%%%%%%%%%%%%%%%%%%%%%%%%%%%%%%%%%%%%%%%%%%%%%%
%%%%%%%%%%%%%%%%%%%%%%%%%%%%%%%%%%%%%%%%%%%%%%%%%%

%%%%%%%%%%%%%%%%%%%%%%%%%%%%%%%%%%%%%%%%%%%%%%%%%%
%%%%%%%%%%%%%%%%%%%%%%%%%%%%%%%%%%%%%%%%%%%%%%%%%%
% Fig.~9.
\begin{figure}
\centering
\includegraphics[width=1\linewidth]{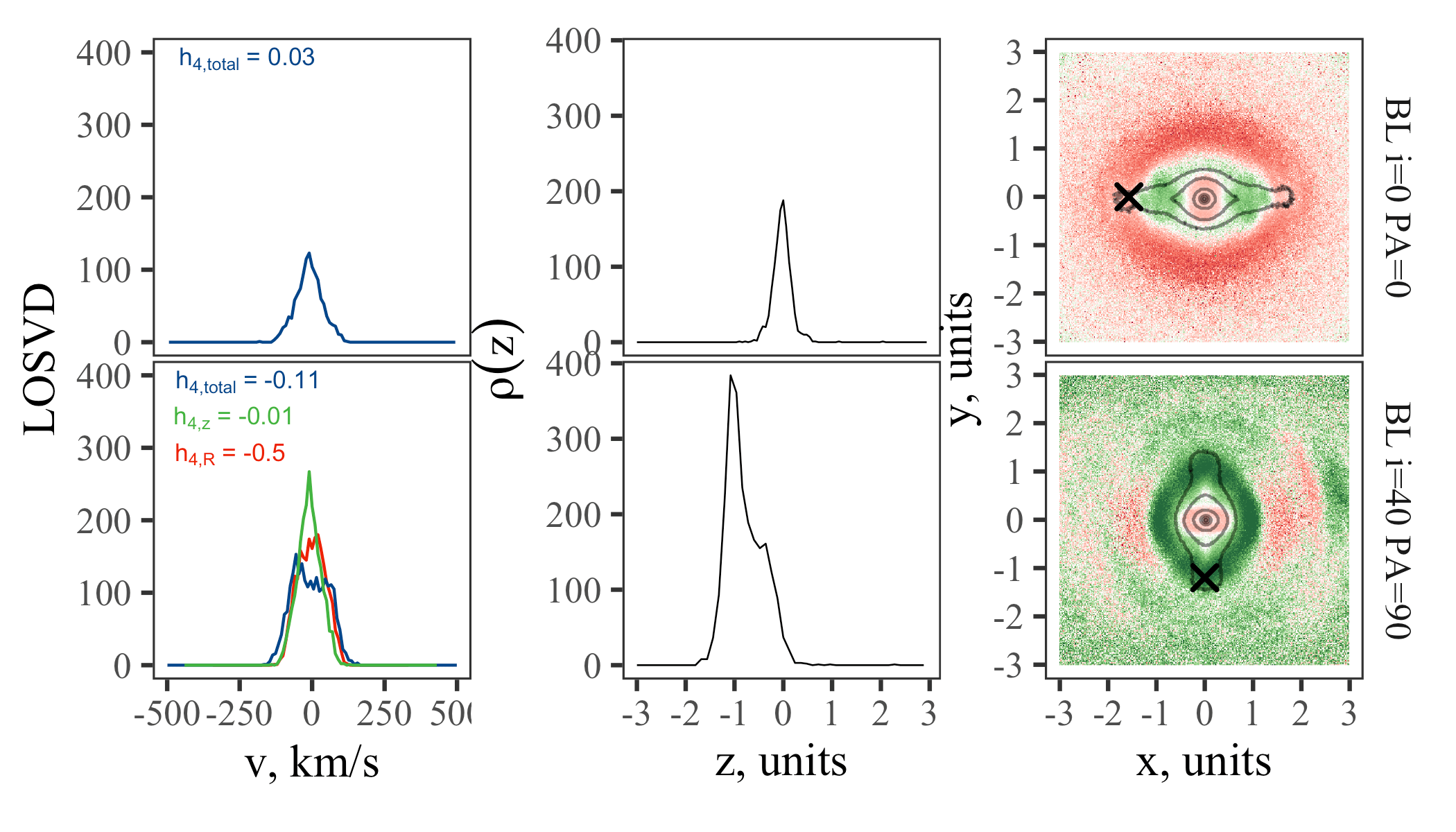}
\caption{\textit{Left:} LOSVDs for the pixel $(-1.2; 0)$ of the BL model at face-on position~(top) and for the same model but inclined by $i=40^{\circ}$ and rotated up to $\mathrm{PA}=90^\circ$~(bottom).  The total LOSV (dark blue color) was additionally decomposed  into vertical~($v_z$ , green color) and radial ($v_{R}$, red) components.
The selected pixel corresponds to the $h_4$ minima along the major axis for the BL model at $i=40^\circ$, $\mathrm{PA}=90^\circ$. \textit{Centre:} the vertical density distribution for both cases. \textit{Right:} $h_4$ maps for $i=0^{\circ}$~(top) and $i=40^\circ$, $\mathrm{PA}=90^\circ$~(bottom). The isophotes~(black lines) correspond to the intensity maps for a given inclination. The cross symbols show the position of selected pixels.}
\label{fig:h4_velcontrib_pa90}
\end{figure}
%%%%%%%%%%%%%%%%%%%%%%%%%%%%%%%%%%%%%%%%%%%%%%%%%%
%%%%%%%%%%%%%%%%%%%%%%%%%%%%%%%%%%%%%%%%%%%%%%%%%%

%%%%%%%%%%%%%%%%%%%%%%%%%%%%%%%%%%%%%%%%%%%%%%%%%%
%%%%%%%%%%%%%%%%%%%%%%%%%%%%%%%%%%%%%%%%%%%%%%%%%%
Our second example is the contamination of the LOSVD by the radial velocity components. For the BL model, we choose the pixel at the bar major axis at the distance of 1.57 units from the centre. Fig.~\ref{fig:h4_alongmajor_all_incpa0} (upper right plot, solid line) and Fig.~\ref{fig:h4_velcontrib_pa90} (upper right plot, a cross in a red area) demonstrate a small positive value of $h_4$ in this pixel. The distribution function $f(v_z)$ in this pixel has a narrow peak around zero (Fig.~\ref{fig:h4_velcontrib_pa90}, upper left plot) and corresponds to a narrow density distribution in the vertical direction (Fig.~\ref{fig:h4_velcontrib_pa90}, upper middle plot) since the line of sight passes exclusively through the peripheral rather thin parts of the bar (Fig.~\ref{fig:BL_layers_los}, right plot, light grey line).
\par
The inclination of the stellar disc around the minor axis of the bar by $i=40^\circ$ shifts the positions along the major axis of the bar and reduces the distances due to projection effects. Thus, the selected pixel end up at the distance of 1.2 units from the centre and falls into a deep negative minimum of $h_4$ (Fig.~\ref{fig:h4_alongmajor_all_incpa0}, right column, solid line, $i=40^\circ$ and Fig.~\ref{fig:h4_velcontrib_pa90}, bottom right plot, a cross in a dark green area). Now, two velocity components, $v_\mathrm{R}$ and $v_z$ contribute to the LOSVD (Fig.~\ref{fig:h4_velcontrib_pa90}, bottom left plot), and the features of the LOSVD  are determined by the features of the radial velocity component distribution function $f(v_\mathrm{R})$. At the same time, the distribution function $f(v_z)$ remains narrow, although the line of sight in the upper half-plane of the B/PS bulge passes partially through its loose parts (Fig.~\ref{fig:BL_layers_los}, right plot, light green line). This is reflected in the skewed density distribution in the vertical direction (Fig.~\ref{fig:h4_velcontrib_pa90}, bottom middle plot). Such a distribution has a negative value of the parameter $d_4$ (Fig.~\ref{fig:h4_alongmajor_all_incpa0}, right column, dashed line, $i=40^\circ$). Additional particles from the loose region have almost no effect on $f(v_z)$, but they broaden the distribution $f(v_\mathrm{R})$ since particles located in a wide range of distances from the centre make an additional contribution to it.
\par
At $i=40^\circ$ and $\mathrm{PA}=90^\circ$, the B/PS bulge itself is located closer to the centre from the negative minimum of $h_4$ along the major axis. 
In this region, there are many negative pixels of $h_4$ and $d_4$ for $i=0^\circ$ which implies that the kinematic diagnostics of peanuts still works here, although the pixels themselves are not located at the very minimum of $h_4$. Thus, at $\mathrm{PA}=90^\circ$ up to inclination $i=40^\circ$, if we take a cut along the major axis of the bar we can assume that the negative minimum of $h_4$ points to the upper boundary of the B/PS bulge along the major axis (its tip).
\par
For $i=60^\circ$ and $\mathrm{PA}=90^\circ$, there is no formal negative minimum of $h_4$ along the major axis in the B/PS bulge region, and the cuts, as well as $h_4$ maps, are difficult to interpret.

%%%%%%%%%%%%%%%%%%%%%%%%%%%%%%%%%%%%%%%%%%%%%%%%%%
%%%%%%%%%%%%%%%%%%%%%%%%%%%%%%%%%%%%%%%%%%%%%%%%%%
\section{Orbital groups and $h_4$}
\label{sec:orbital_groups}
% 5.
%%%%%%%%%%%%%%%%%%%%%%%%%%%%%%%%%%%%%%%%%%%%%%%%%%
%%%%%%%%%%%%%%%%%%%%%%%%%%%%%%%%%%%%%%%%%%%%%%%%%%
In this section, we discuss in detail the features of $h_4$ maps for all our models and contributions of various orbital components to the LOSVD (Section~\ref{sec:families} contains information about the bar dissection into orbital groups). In \citet{Smirnov_etal2021} and \citet{Tikhonenko_etal2021}, each of our models was disassembled into several orbital groups to analyse the face-on and edge-on morphology of an overall bar. 
We have excluded each orbital group in turn from the final model  to determine how the $h_4$ maps change with the changes in the orbital groups that constitute the bar.
These results should be interpreted with caution since such a simple removal of a particular orbital group does not eliminate the gravitational potential of these points in the rest of the model. This means that the influence of removed particles would be still imprinted in the kinematics of remaining particles and will show up in some way on the maps. \footnote{As an example of such a pitfall, one may try to naively remove the thickest layers of the B/PS bulge only to find that face-on $h_4$ maps still look almost the same as they were for the original model.}
\par
For the face-on models, we analyse which orbital groups are responsible for peanut signatures on the $h_4$ maps. We now focus only on the BL and X models for a better example since the existence of the big bulges blurs the picture for the BLx and Xb models,
albeit the reasons for the appearance of certain features on the $h_4$ maps are the same for all four models.
The corresponding maps for the BLx and Xb models can be found in Appendix~\ref{app:blx_xb_orbits}.

%%%%%%%%%%%%%%%%%%%%%%%%%%%%%%%%%%%%%%%%%%%%%%%%%%
%%%%%%%%%%%%%%%%%%%%%%%%%%%%%%%%%%%%%%%%%%%%%%%%%%
\subsection{Exclusion of orbital groups}
\label{sec:h4_exlusion_faceon}
% 5.1
%%%%%%%%%%%%%%%%%%%%%%%%%%%%%%%%%%%%%%%%%%%%%%%%%%
%%%%%%%%%%%%%%%%%%%%%%%%%%%%%%%%%%%%%%%%%%%%%%%%%%

\begin{figure*}
\centering
\includegraphics[width=1\linewidth]{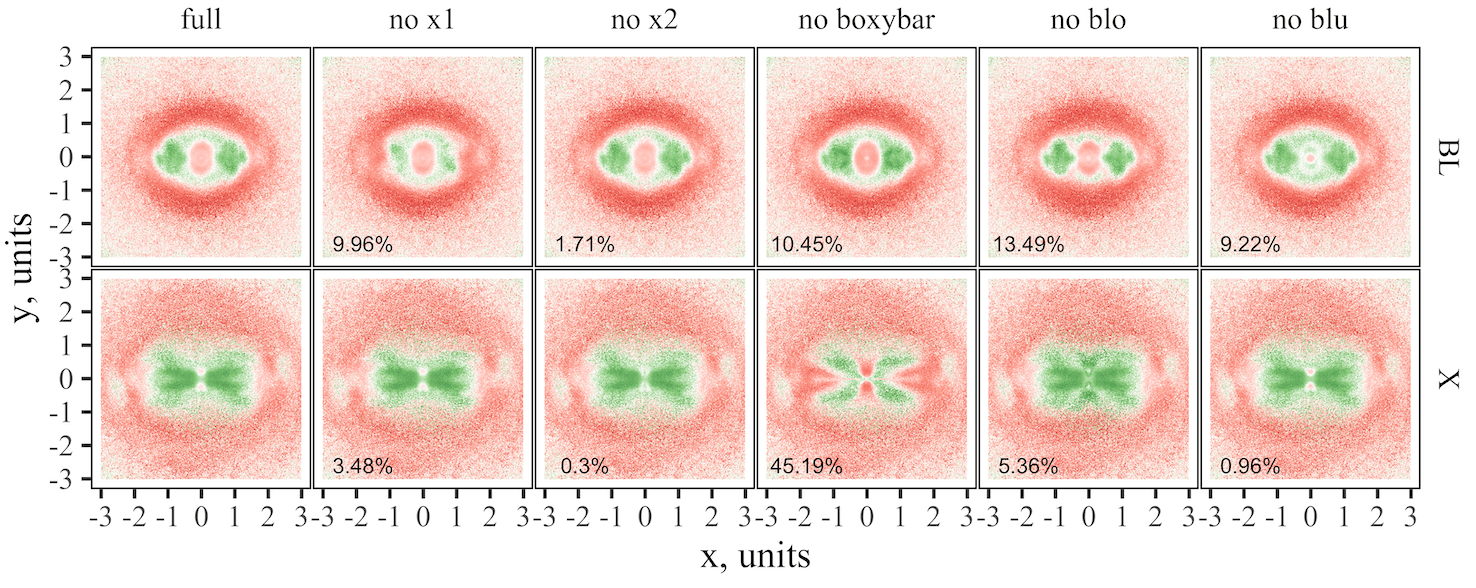}
\caption{The maps of $h_4$ for the BL (\emph{top} row) and X (\emph{bottom} row) models at face-on position with orbital groups excluded in turn. The numbers in the bottom left corner of each plot correspond to the relative contribution of the currently excluded orbital group (indicated in the title of each plot) to the total number of particles~(see Table.~\ref{tab:familiesnumbers} for details). The colors are the same as in the Fig.~\ref{fig:2Dmaps_i0PA0}.}
\label{fig:d4_exclude_orbits}
\end{figure*}

Fig.~\ref{fig:d4_exclude_orbits} shows how the $h_4$ maps change when the orbits of each orbital group are excluded from the face-on snapshots for the BL and X models.
\par
The BL model (upper row) consists of 55.49\% particles of the disc, \textcolor{black}{while around 45\% belong to the bar component.}
%bar particles as the rest. 
The contributions of orbits supporting the bar are approximately equal between x1-like orbits (responsible for a narrow and elongated bar), boxy orbits (creating a face-on peanut), and \blo and \blu orbits~(main components of a barlens,  \citealp{Smirnov_etal2021}) --- each group contributes about~10\% of orbits (Table.~\ref{tab:familiesnumbers}). Although the \blo and \blu orbits are the largest groups of orbits belonging to the bar and supporting the central barlens in the BL model, the exclusion of these types of orbits does not change the main features of the kinematic maps of $h_4$. 
We only note changes on the minor axis, where the number of pixels with negative values decreased after the elimination of the \blo orbits.
On the other side, when the x1 orbits are removed, almost all pixels with negative values of $h_4$ on the major axis disappear ~(upper row, the 2nd column of Fig.~\ref{fig:d4_exclude_orbits}).
Thus,  the characteristic minima of $h_4$ should be associated with the narrow and elongated bar and not with the barlens for the BL model. 
This result is consistent with the conclusions by \citet{Tikhonenko_etal2021} obtained by analysing snapshots of various orbital subsystems, but here we use the kinematic analysis.
\par
In its turn, the model with a pure face-on peanut bar~(without a classical bulge and without a barlens, labelled X) mainly consists of disc~(44.71\%) and boxy orbits~(45.19\%). Accordingly, the elimination of boxy orbits for the X model removes almost all kinematic features in the centre of the model galaxy, along with four dark green rays with a small opening angle. 
All other groups of orbits in the bar are not represented in large numbers~(only about 10\% in total), and their exclusion has no effect on the $h_4$ maps.
\par
For the BL model, a signal from pixels with negative values of $d_4$ is associated almost exclusively with x1 orbits. For the X model, on the contrary, a characteristic pattern in the area of the bar with negative values of the parameter $d_4$ is created by a subsystem assembled from boxy orbits. Thus, for models with different bar morphology, the features of the LOSVD expressed in term of $h_4$ parameter are formed by different orbits. 

%%%%%%%%%%%%%%%%%%%%%%%%%%%%%%%%%%%%%%%%%%%%%%%%%%
%%%%%%%%%%%%%%%%%%%%%%%%%%%%%%%%%%%%%%%%%%%%%%%%%%
% Fig.~10
%%%%%%%%%%%%%%%%%%%%%%%%%%%%%%%%%%%%%%%%%%%%%%%%%%
%%%%%%%%%%%%%%%%%%%%%%%%%%%%%%%%%%%%%%%%%%%%%%%%%%

%%%%%%%%%%%%%%%%%%%%%%%%%%%%%%%%%%%%%%%%%%%%%%%%%%
%%%%%%%%%%%%%%%%%%%%%%%%%%%%%%%%%%%%%%%%%%%%%%%%%%
\subsection{Exclusion of orbital groups from inclined models}
\label{sec:orbit_exl_4090}
% 5.2
%%%%%%%%%%%%%%%%%%%%%%%%%%%%%%%%%%%%%%%%%%%%%%%%%%
%%%%%%%%%%%%%%%%%%%%%%%%%%%%%%%%%%%%%%%%%%%%%%%%%%
In this section, we focus on a more detailed discussion of the formation of the ring of $h_4$ minima for barlens galaxies. 
As was discussed in previous sections, the ring of $h_4$ minima is not characteristic feature of galaxies with barlenses, and this ring is formed for all our models due to the contribution of azimuthal and radial velocity components to the LOSVD.
In order to demonstrate that the pixels with negative values of $h_4$ on a face-on maps for inclined models have different nature, we now focus on the $h_4$ maps when the model has inclination $i=40^\circ$ and the position angle $\mathrm{PA}=90^\circ$~(the most prominent ring). We will exclude the same orbital groups as in Section~\ref{sec:h4_exlusion_faceon} from the model.
\par
\begin{figure*}
\centering
\includegraphics[width=1\linewidth]{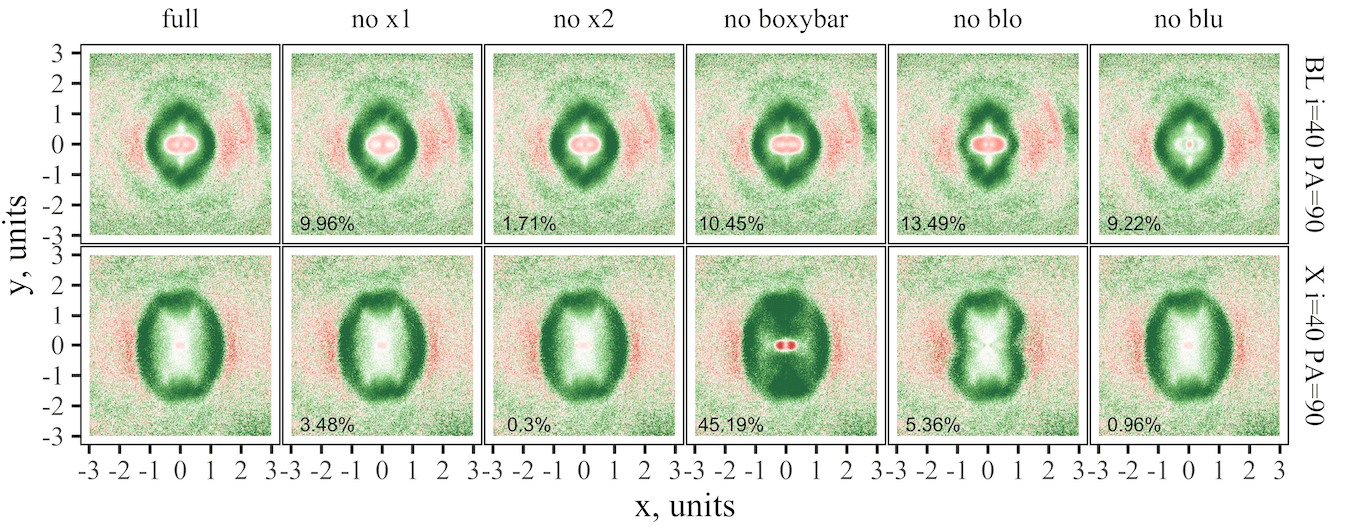}
\caption{The same maps of $h_4$ for the for BL (\emph{top} row) and X (\emph{bottom} row) models as in Fig.~\ref{fig:d4_exclude_orbits}  but for $i=40^\circ$ and $\mathrm{PA}=90^\circ$. Starting from the second column, a particular orbital group is excluded in each panel. The colors are the same as in Fig.~\ref{fig:2Dmaps_i0PA0}.}
\label{fig:h4_exclude_orbits_4090}
\end{figure*}

Fig.~\ref{fig:h4_exclude_orbits_4090}~(top row) shows the results for the BL model. 
%for $i=40^\circ$ and $\mathrm{PA}=90^\circ$. 
The figure shows that exclusion of x1 orbits does not change the $h_4$ map\footnote{The same area as for $i=0^\circ$ and $\mathrm{PA}=0^\circ$ case in Section~\ref{sec:h4_exlusion_faceon}.}, although at $i=0^\circ$ and $\mathrm{PA}=0^\circ$ the x1 orbits are responsible for the $h_4$ minima. The exclusion of the \blo orbits changes the map of $h_4$, in contrast to the face-on case, producing some effect along the minor axis, but still not eliminating the ring. 
\par
The bottom row of the Fig.~\ref{fig:h4_exclude_orbits_4090} shows the $h_4$ maps for the inclined X model. 
After eliminating the main orbital groups for the X model~(`boxy' orbits), we also do not observe the elimination of the ring of $h_4$ minima\footnote{Although we note that there are practically no particles in this region.}. Moreover, the exclusion of any type of orbits does not remove the ring. 
\par 
Thus, we associate the ring of $h_4$ minima for $i>20^\circ$ and $\mathrm{PA}>45$ not with the morphological features of the bar of one model or another, but with the contamination of the LOSVD by the velocity components $v_\varphi$ or $v_\mathrm{R}$. 
As was shown in Section~\ref{sec:LOSVD_vel_components}, the main contribution to the LOSVD contamination for cuts along the major axis of the bar comes from the radial velocity component, if the galaxy is inclined around the minor axis of the bar. As to the minor axis, the appearance of the negative pixels comes from the the azimuthal velocity component.
Thus, we attribute the ring of $h_4$ minima for an inclined galaxy to a superposition of these two effects. For pixels in an intermediate position between the major and minor axes, the ring arises due to the contribution of all velocity components to the LOSVD, but not due to the features of the vertical density distribution.
%%%%%%%%%%%%%%%%%%%%%%%%%%%%%%%%%%%%%%%%%%%%%%%%%%
%%%%%%%%%%%%%%%%%%%%%%%%%%%%%%%%%%%%%%%%%%%%%%%%%%
% Fig.~11
%%%%%%%%%%%%%%%%%%%%%%%%%%%%%%%%%%%%%%%%%%%%%%%%%%
%%%%%%%%%%%%%%%%%%%%%%%%%%%%%%%%%%%%%%%%%%%%%%%%%%

%%%%%%%%%%%%%%%%%%%%%%%%%%%%%%%%%%%%%%%%%%%%%%%%%%
%%%%%%%%%%%%%%%%%%%%%%%%%%%%%%%%%%%%%%%%%%%%%%%%%%
\section{Discussion}
\label{sec:discussion}
% 6.
%%%%%%%%%%%%%%%%%%%%%%%%%%%%%%%%%%%%%%%%%%%%%%%%%%
%%%%%%%%%%%%%%%%%%%%%%%%%%%%%%%%%%%%%%%%%%%%%%%%%%
The kinematic maps presented in this work are well suited for interpreting IFU spectroscopy data. To date, one of the best spectral data on the kinematics of galaxies is the TIMER data. The galaxies of this sample (21 objects) have inclinations from $i=20^\circ$ to $i=60^\circ$ and a PA spread from $0^\circ$ to $90^\circ$.  
{Thus, to obtain an interpretation of these data, one needs a grid of $N$-body models at intermediate inclinations. However, \citet{Debattista_etal2005} did not provide any maps with  nonzero inclinations. They only briefly discussed the inclination effects for $i<30^\circ$.}
\citet{Bureau_Athanassoula2005} demonstrated the kinematic cuts, in particular, of the parameter $h_4$ (their Fig.~3), with different inclinations and PA (mainly $\mathrm{PA}=90^\circ$) but only for $i \geq 75^\circ$. \citet{Iannuzzi_Athanassoula2015} also mainly discussed maps for models at high inclinations. Thus, our choice of inclinations  ($0^\circ, 20^\circ, 40^\circ, 60^\circ$) and position angles ($0^\circ, 45^\circ, 90^\circ$) fill in the gap that exists in previous works.
According to \citet{Laurikainen_Salo2017}, 11 from 17 clearly barred galaxies are also barlens galaxies in \cite{Gadotti_etal2020} TIMER sample. One more galaxy (NGC~7755), which is not included in the sample by \citet{Laurikainen_Salo2017}, has clear photometric signs of a barlens on the S$^4$G image. To interpret such kinematic data, it is very important to have numerical models with a barlens. The models by \citet{Debattista_etal2005} and \citet{Bureau_Athanassoula2005} do not include a pure barlens. 
\citet{Iannuzzi_Athanassoula2015} have barlens models~(models GTR101 and GTR116), but there is no detailed map grid for such galaxies at intermediate inclinations. We provide these maps for our BL model with a barlens and demonstrate the kinematic features associated with this subsystem. Moreover, our wide grid of inclinations and position angles for all our models provide the possibilities to interpret not only MUSE data but also apply it for future spectroscopic surveys.
\par
We have models that are close to the models analysed by \citet{Debattista_etal2005}, and using their example we can confirm the absence of a relationship between the extent of the B/PS bulge and the size of the area where the double negative minima of $h_4$ are localised. Our model X resembles the R1 and R4 models by \citet{Debattista_etal2005} in the Toomre parameter $Q$ and the initial thickness, although \citet{Debattista_etal2005} considered models with rigid halos. The Xb model is analogous to the model B3 ($M_\mathrm{b}/M_\mathrm{d}=0.2$, \citealp{Debattista_etal2005}). As a result, the $h_4$ maps for the models X, Xb are similar to the maps of the parameter $s_4$ of \citet{Debattista_etal2005} for their models R1, R4 and B3. The X model has the extended B/PS bulge, and the negative minima of $h_4$ are located close to the center and do not coincide with the thickest parts of the B/PS bulge, as it is seen on the side-on snapshots (Fig.~\ref{fig:models_view}). To some extent, this also applies to the Xb model. As was noted in the Section~\ref{sec:LOSVD}, the negative values of the $h_4$ parameter are also partially localised in the bar itself away the minima on the major axis along four rays and outline the dimensions of the B/PS bulge. The models by \citet{Debattista_etal2005}, similar to the X and Xb models, exhibit the same behavior (e.g., the rightmost panel of their Fig.~13). This means that although the B/PS bulge is usually associated with the thickest parts of the bar, the parameter $h_4$ does not always delineate the boundaries of these thickest areas if we take cuts along the major axis. In Section~\ref{sec:discussion_1} we will discuss in more details the relationship between the bar thickness, vertical density distribution, and kinematic signatures of a peanut.
\par
One more feature on the $h_4$ maps, which is typical for all our models is the ring of negative values of $h_4$ when $\mathrm{PA}> 45^\circ$. Signatures of a ring from negative values of $h_4$ can be noticed in figure~29 by \citet{Iannuzzi_Athanassoula2015} (the case $\mathrm{PA}=90^\circ$, $i=60^\circ$, models GTR101 and GTR116, which apparently have barlenses). This is the main plot which is used to interpret the TIMER data. As the appearance of a ring on the $h_4$ maps for our models is solely the effect of the galaxy inclination and a special value of the position angle of the bar major axis, in Section~\ref{sec:discussion_2} we discuss the pitfalls regarding the interpretation of available observational kinematic maps.
We also warn of difficulties regarding the interpretation of available observational kinematic maps.
%%%%%%%%%%%%%%%%%%%%%%%%%%%%%%%%%%%%%%%%%%%%%%%%%%
%%%%%%%%%%%%%%%%%%%%%%%%%%%%%%%%%%%%%%%%%%%%%%%%%%

\subsection{How kinematics captures the density distribution}
\label{sec:discussion_1}
\begin{figure}
\centering
\includegraphics[width=1\linewidth]{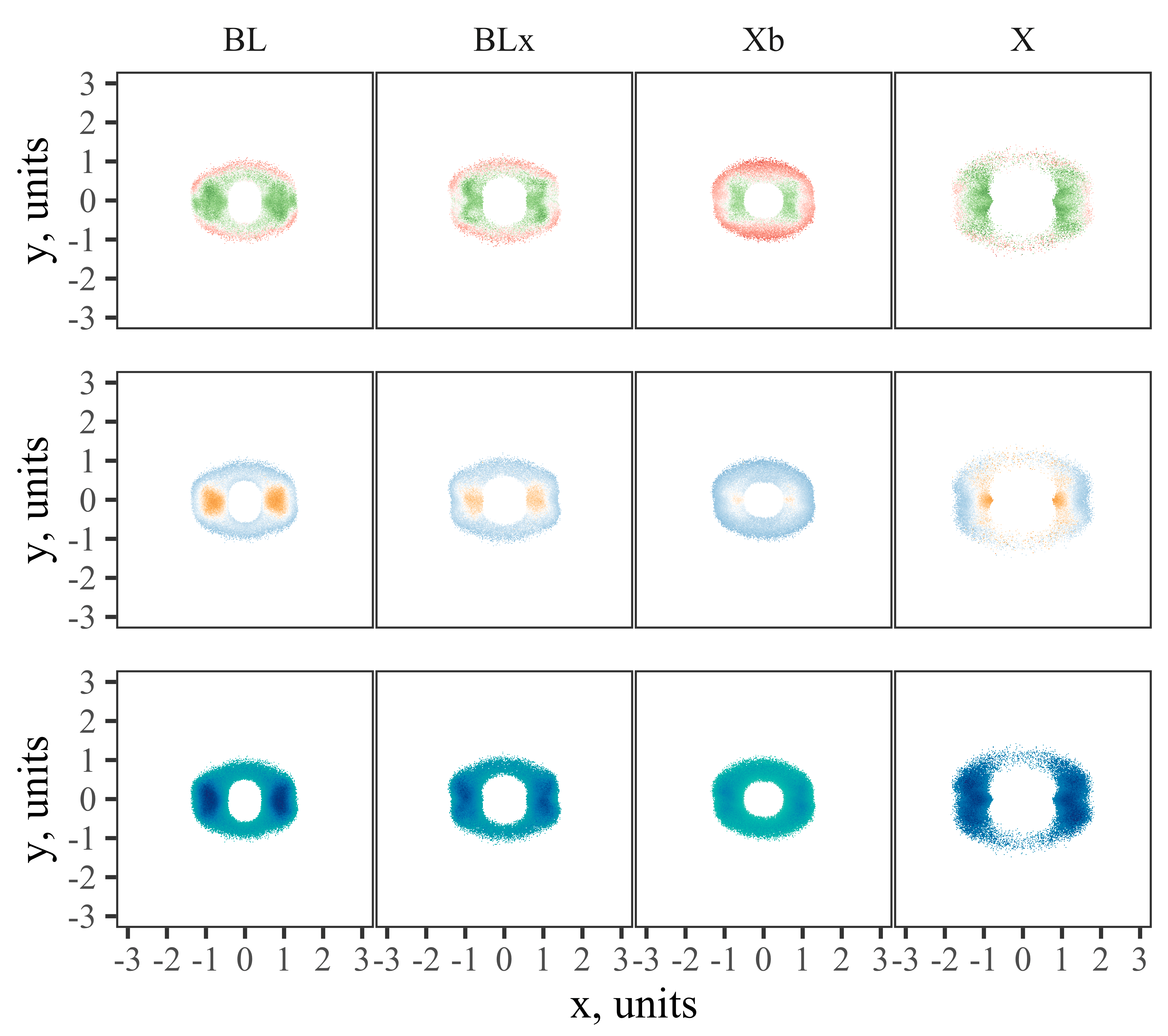}
\caption{The maps of $h_4$~(\emph{top} row), $d_4$~(\emph{middle} row) and $\med|z|$~(\emph{bottom} row) for the BL, BLx, Xb, and X models (from the left to the right) in the thickest area of the bar~(top 10\% of the  pixels  sorted by $\med|z|$). The colors are the same as in Fig.~\ref{fig:2Dmaps_i0PA0}.}
\label{fig:h4_in_thicker0.7_places}
\end{figure}

\begin{figure*}
\centering
\includegraphics[width=0.48\linewidth]{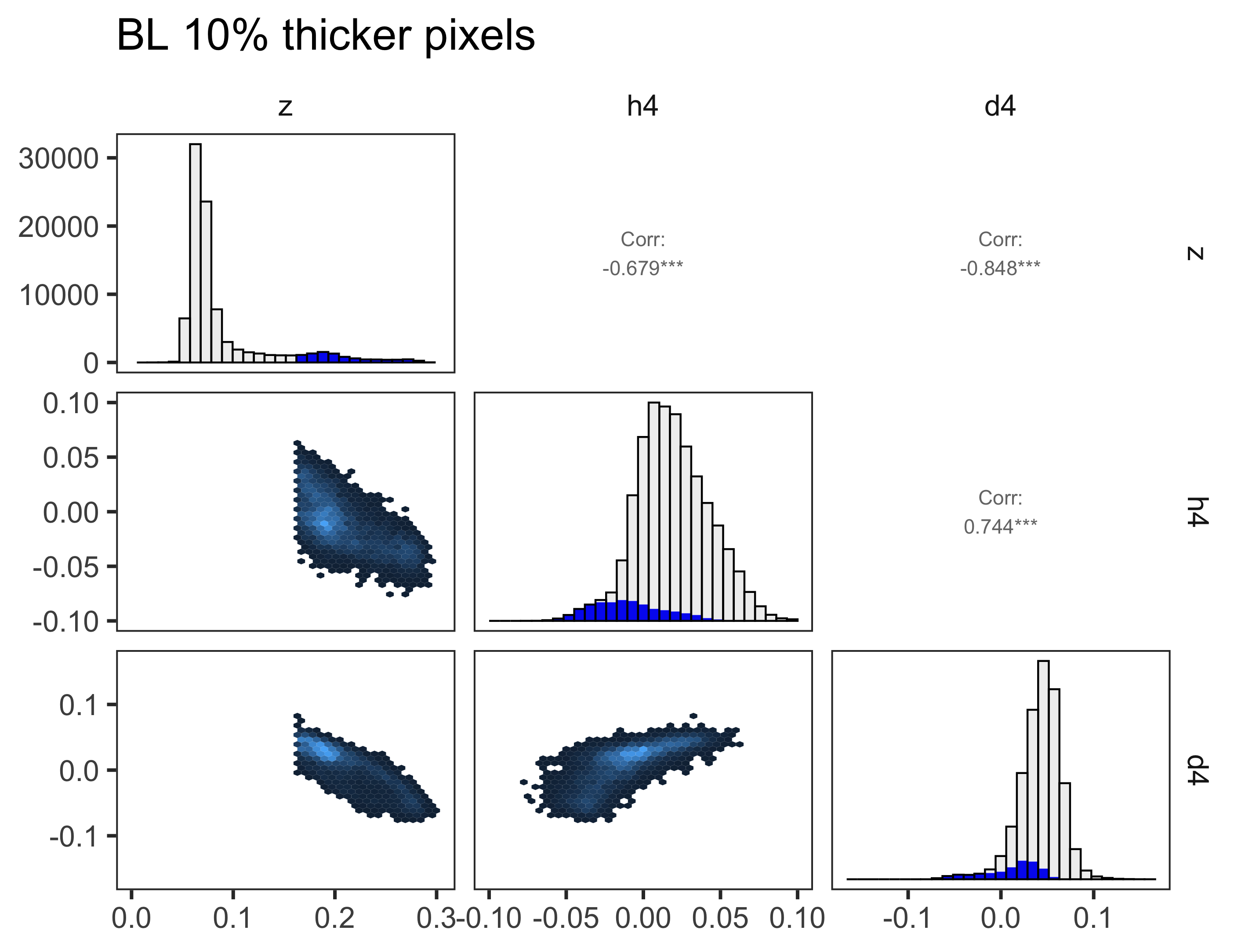}
\includegraphics[width=0.48\linewidth]{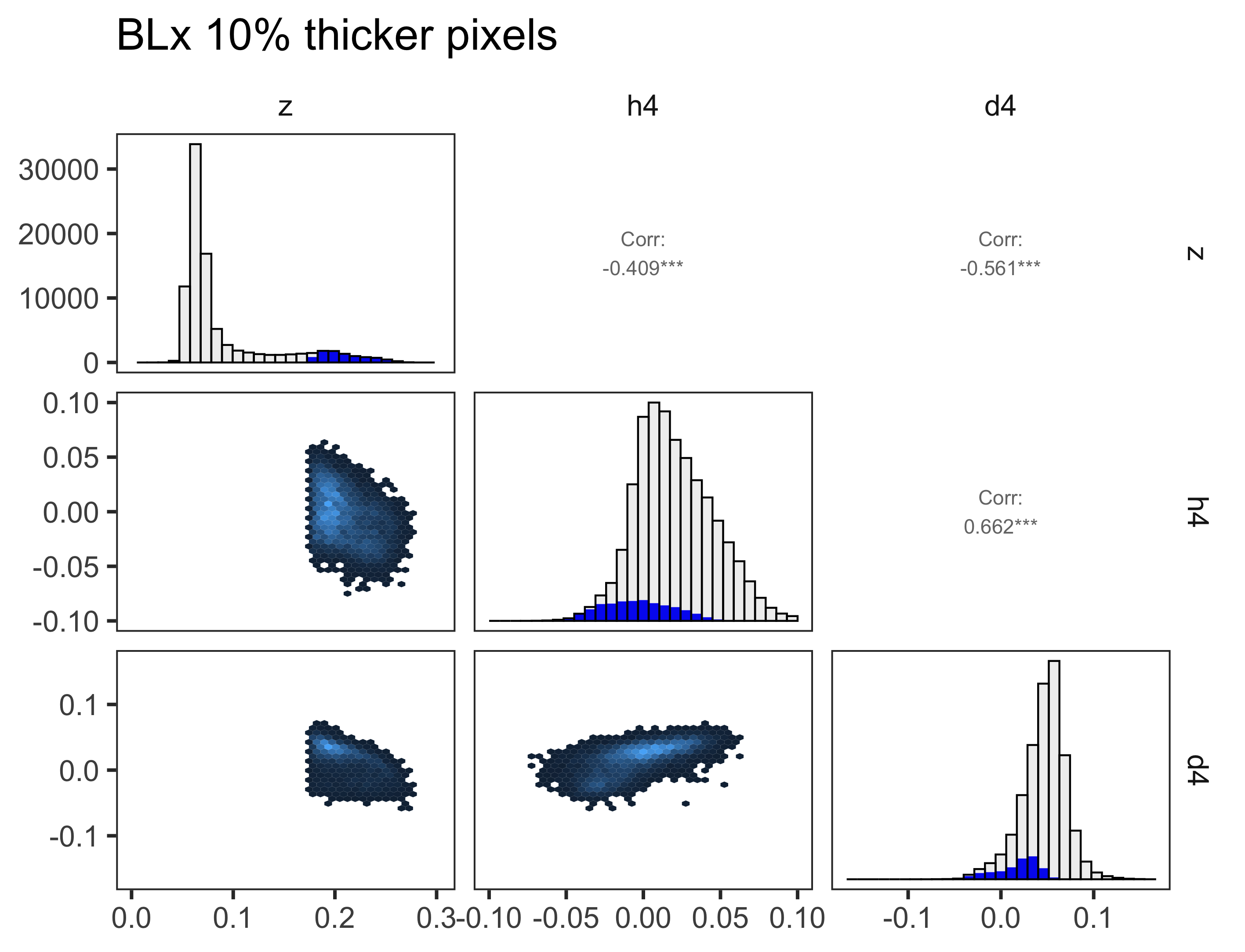}
\includegraphics[width=0.48\linewidth]{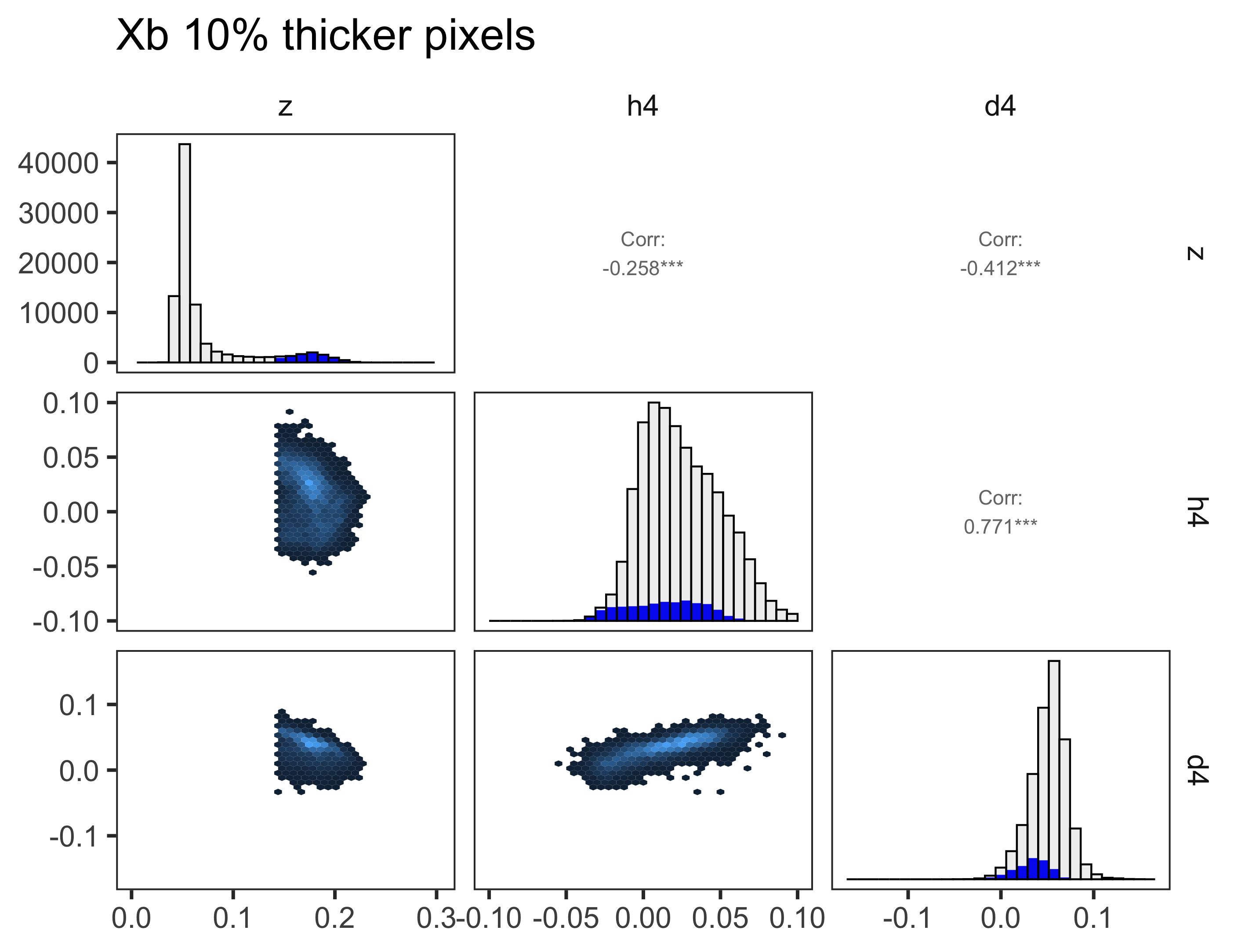}
\includegraphics[width=0.48\linewidth]{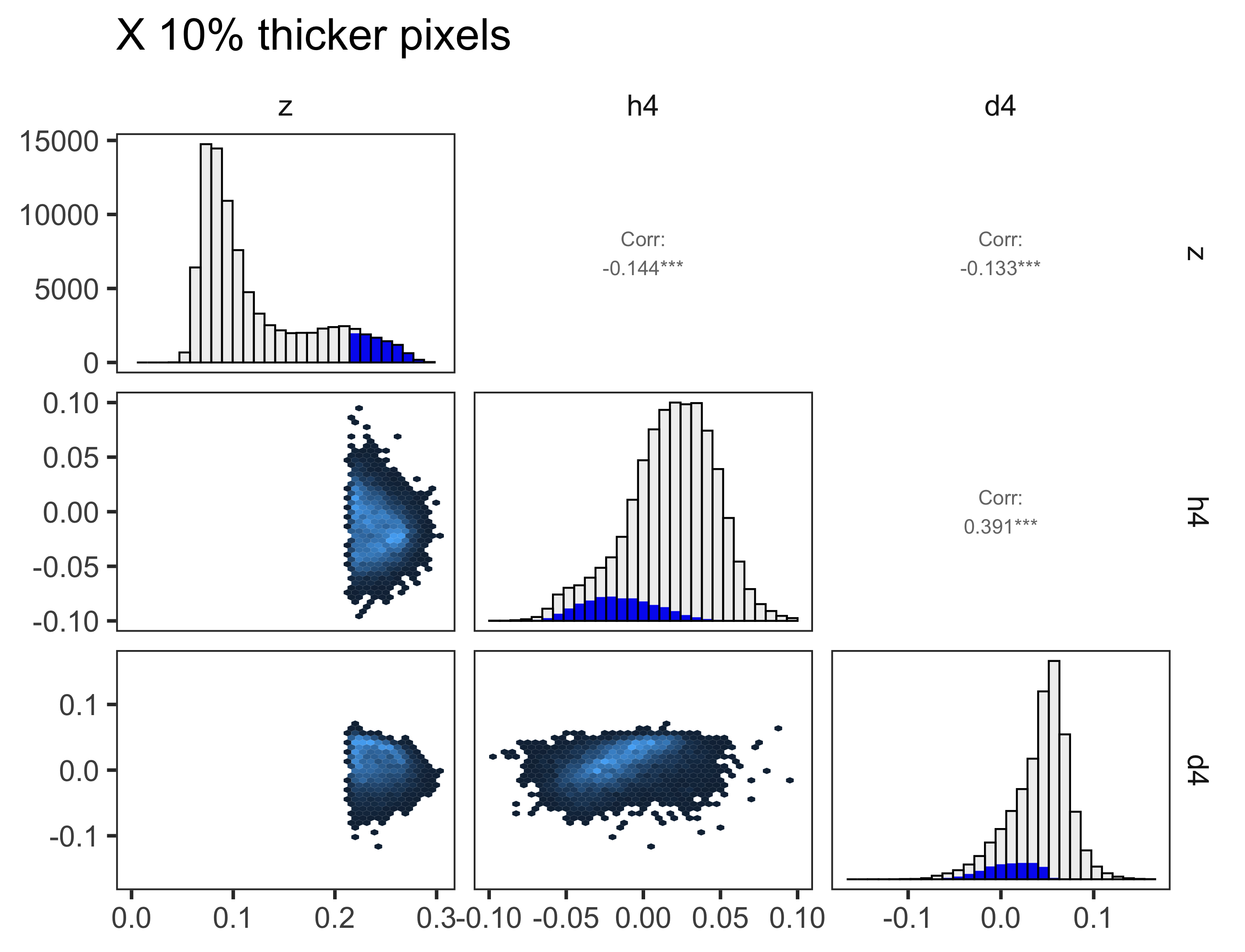}
\caption{The correlation matrix between $h_4$, $d_4$, and $\med|z|$ for the thickest areas of the bar~(top 10\% of the pixels sorted by $\med|z|$), according to Fig.~\ref{fig:h4_in_thicker0.7_places}}
\label{fig:h4d4medz_corr}
\end{figure*}

% 6.1
%%%%%%%%%%%%%%%%%%%%%%%%%%%%%%%%%%%%%%%%%%%%%%%%%%
%%%%%%%%%%%%%%%%%%%%%%%%%%%%%%%%%%%%%%%%%%%%%%%%%% 
In Section~\ref{sec:LOSVD_i0}, we have shown that for the models with B/PS bulges but with different bar morphology in a face-on view, the characteristic vertical density distribution of B/PS bulges~(the values of $d_4$) is followed by the broadness of LOSVD~($h_4$ parameter).
This result is in agreement with \cite{Debattista_etal2005}. Moreover, we have shown that negative values of $h_4$ follow only flat-topped distribution, but not the thickest parts (Fig.~\ref{fig:2Dmaps_i0PA0}). For a more detailed analysis, we now discuss the behaviour of $h_4$ and $d_4$ values in the thickest areas of the bar~(top 10\% of the pixels  sorted by $\med|z|$).
Fig.~\ref{fig:h4_in_thicker0.7_places} shows that the thickest part of the galaxy is not characterised by a flat-topped density distribution and, respectively, by negative values of $h_4$.
Also, Fig.~\ref{fig:h4_in_thicker0.7_places} demonstrates that in the thickest regions of the bar, the most negative values of $h_4$ and $d_4$ are localised in the same places for all models. At least, this is true for pixels lying in a strip along the major axis of the bar. 
At the same time, there is no connection between the kinematics ($h_4$, the top row of Fig.~\ref{fig:h4_in_thicker0.7_places}) and the thickness of the models~(bottom row of Fig.~\ref{fig:h4_in_thicker0.7_places}). All considered models show both positive and negative $h_4$ values in these selected pixels.
The $h_4$ minima do not fall into the regions of the largest thickness for the BLx and Xb models, especially for the X model, although for the BL model, the $h_4$ minima are located in the pixels that characterise the largest thickness. Fig.~\ref{fig:h4d4medz_corr} shows the relationships between three parameters, $h_4$, $d_4$ and $\med|z|$, for the pixels in the thickest areas of the bar (10\% of the pixels with the largest thickness values). In the thickest zones, a clear anticorrelation between $h_4$ and $\med|z|$ exists only for the BL model. 
\par
For the thickest parts of the bar in models without a barlens, the minima of $h_4$ and $d_4$ shift from the centre to the periphery (compare Fig.~\ref{fig:2Dmaps_i0PA0} and Fig.~\ref{fig:h4_in_thicker0.7_places}).
In the X model, the prevailing orbits in the bar are boxy orbits, which, unlike x1 orbits, have many self-intersections near the centre. They give deep two-sided minima of $h_4$ near the centre on the maps in Fig.~\ref{fig:2Dmaps_i0PA0}, although these areas are not thickest. For the thickest areas of the bar, the positions of the double minima of $h_4$ and $d_4$ are very similar for all models. We refer it to the fact that the minima for our X model are very wide and the negative values of $h_4$ are still present within boundary of the most vertically extended B/PS bulge part (see Fig.~\ref{fig:models_view}). 
\par
Thus, although the flat-topped shape of the LOSVD~(negative values of $h_4$) in the face-on case for models with different types of B/PS bulges does not unambiguously track the thickest areas of the B/PS bulges, in the thickest regions, the $h_4$ parameter certainly is an indicator of the features of the vertical density distribution~($d_4$). It follows that interpretation of the real data has to be done carefully since the minima of $h_4$ do not track the position of the thickest zones and, accordingly, do not track the size of the B/PS bulge.

%%%%%%%%%%%%%%%%%%%%%%%%%%%%%%%%%%%%%%%%%%%%%%%%%%
%%%%%%%%%%%%%%%%%%%%%%%%%%%%%%%%%%%%%%%%%%%%%%%%%%
% Fig.~12
%%%%%%%%%%%%%%%%%%%%%%%%%%%%%%%%%%%%%%%%%%%%%%%%%%
%%%%%%%%%%%%%%%%%%%%%%%%%%%%%%%%%%%%%%%%%%%%%%%%%%

%%%%%%%%%%%%%%%%%%%%%%%%%%%%%%%%%%%%%%%%%%%%%%%%%%
%%%%%%%%%%%%%%%%%%%%%%%%%%%%%%%%%%%%%%%%%%%%%%%%%%
% Fig.~13
%%%%%%%%%%%%%%%%%%%%%%%%%%%%%%%%%%%%%%%%%%%%%%%%%%
%%%%%%%%%%%%%%%%%%%%%%%%%%%%%%%%%%%%%%%%%%%%%%%%%%
\par
The $h_4$ parameter reliably tracks the vertical density distribution only in the face-on case. As the inclination increases, other velocity components ($v_{\varphi}$ and $v_{R}$) contaminate the LOSVD, and $h_4$ no longer follows $d_4$. Thus, for inclined galaxies, the analysis of the shape of the LOSVD as an indicator of a peanut should be carried out with caution because at $i\geq40^\circ$ the connection between $h_4$ and $d_4$ is broken (Fig.~\ref{fig:h4_alongmajor_all_incpa0}). There is another effect associated with the broadness of LOSVD. Differently rotating components (for example, disc and bar) also lead to a negative value of the parameter $h_4$, but this has nothing to do with the features of the vertical density distribution. Such effects also may contaminate the $h_4$ maps of inclined galaxies. We have shown that for all our models with B/PS bulges, the effects of inclination and rotation of the bar major axis affect the $h_4$ maps similarly.  Accordingly, we conclude that there are no significant differences in kinematics between different types of B/PS bulges in our models.

%%%%%%%%%%%%%%%%%%%%%%%%%%%%%%%%%%%%%%%%%%%%%%%%%%
%%%%%%%%%%%%%%%%%%%%%%%%%%%%%%%%%%%%%%%%%%%%%%%%%%
\subsection{Possible pitfalls in interpreting observational data}
\label{sec:discussion_2}
% 6.2
%%%%%%%%%%%%%%%%%%%%%%%%%%%%%%%%%%%%%%%%%%%%%%%%%%
%%%%%%%%%%%%%%%%%%%%%%%%%%%%%%%%%%%%%%%%%%%%%%%%%%
The $h_4$ parameter reliably tracks the vertical density distribution only in the face-on case. However, even for galaxies with low inclinations and low PA~(as in the two upper plots on the left of Fig.~\ref{fig:BL_all_i} and Fig.~\ref{fig:h4_alongmajor_all_incpa0}), interpreting the two-sided minima of $h_4$ along the major axis of a bar is sophisticated. The reasonable question is, which component of the bar these minima can be attributed to as the signatures of a B/PS bulge, i.~e. to the bar itself or to the barlens if each of them is present, as in galaxy NGC~98 \citep{Mendez-Abreu_etal2008,Mendez-Abreu_etal2014}. The analysis of the orbital composition of the bar carried out in Section~\ref{sec:h4_exlusion_faceon} showed that for the BL model, only the component assembled from the x1 orbits shows the peanut signatures. Inclusion in or exclusion from the model of orbits assembled into a barlens (\blo and \blu) has no effect on the $h_4$ maps (Fig.~\ref{fig:d4_exclude_orbits}). Signatures of a peanut, as a two-sided minima of $h_4$ on the major axis of the bar, are always present, but they are not related to the barlens.
\par
\begin{figure}
\centering
\includegraphics[width=1\linewidth]{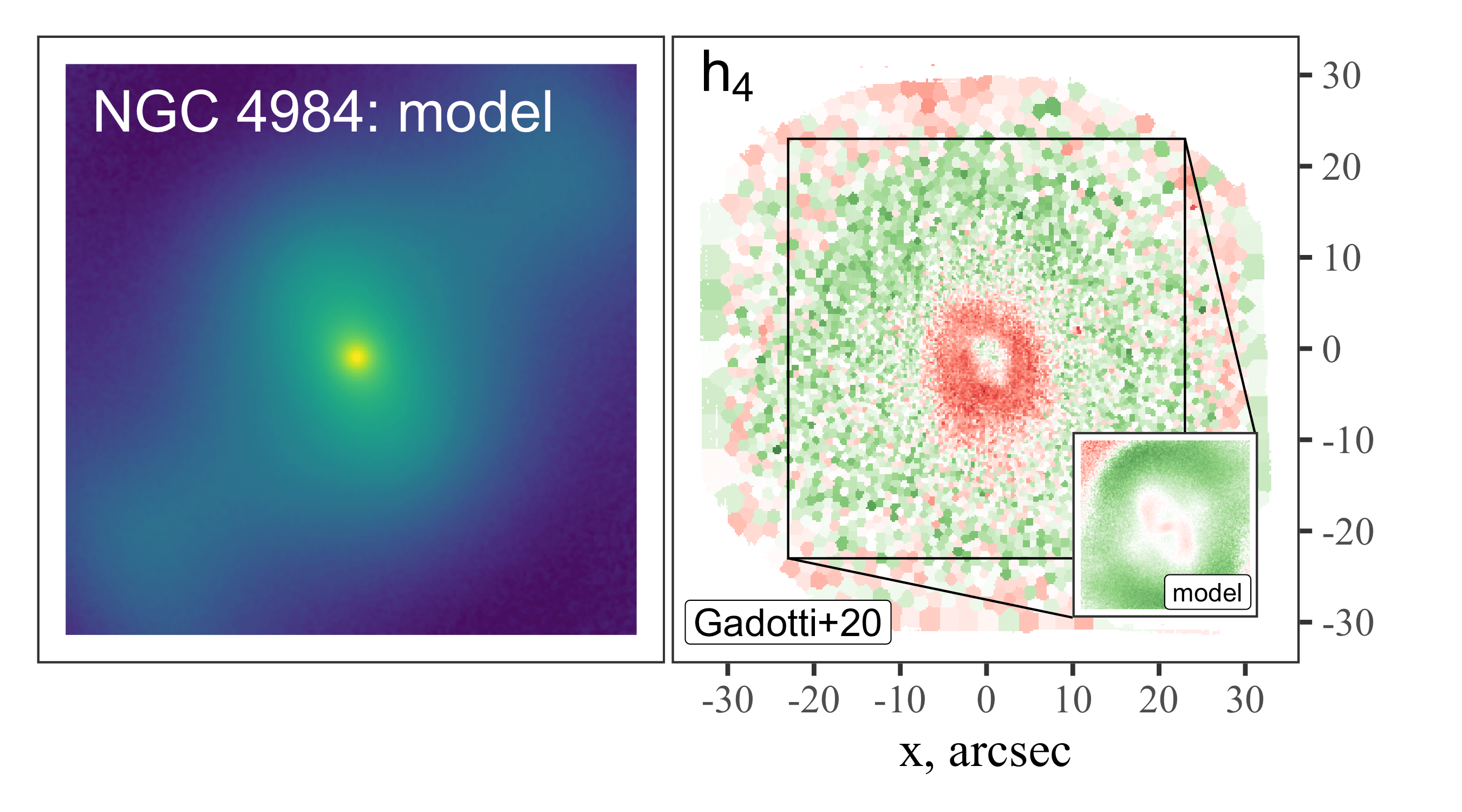}
\includegraphics[width=1\linewidth]{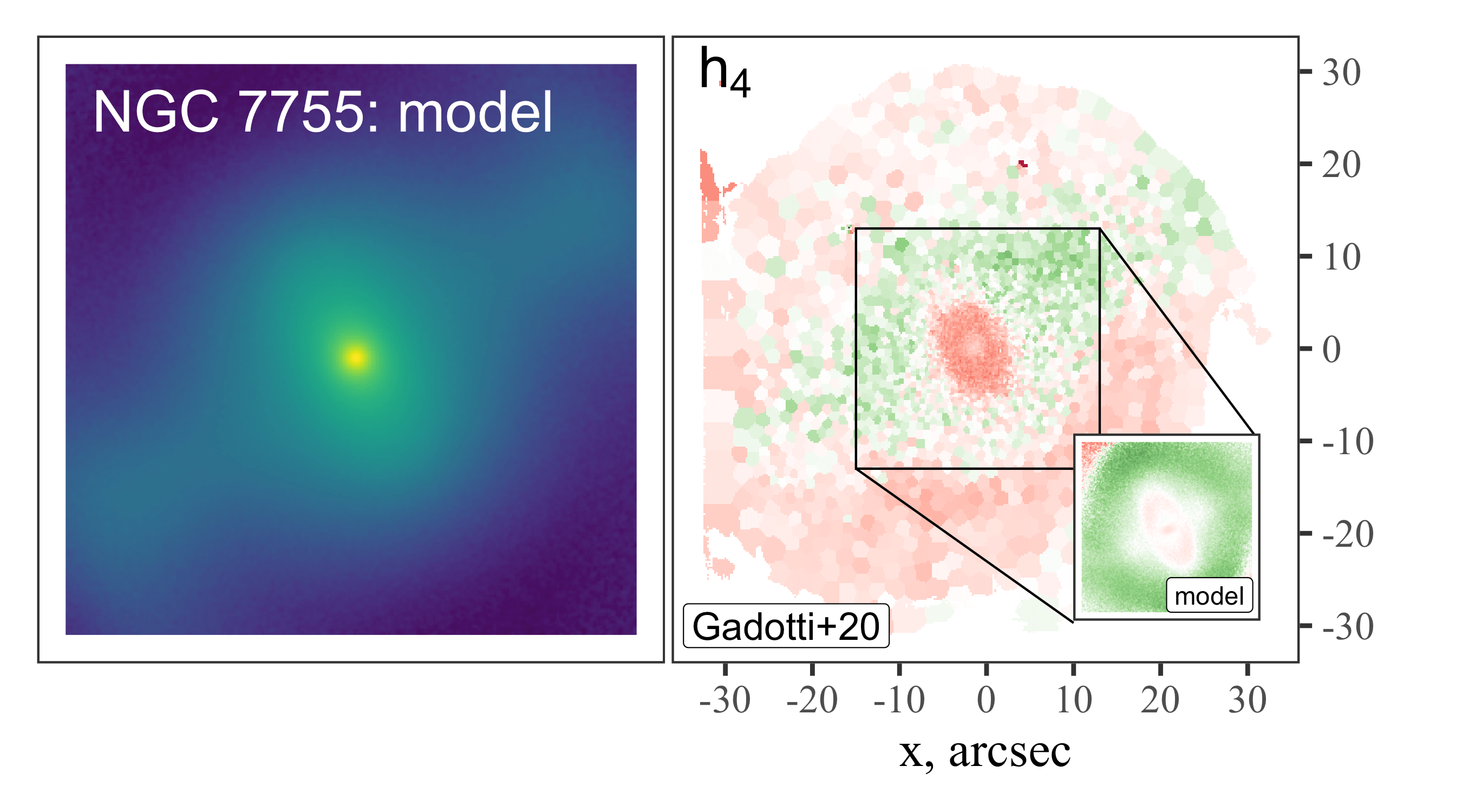}
\caption{
The BL model rotated to mimic two galaxies from the TIMER sample: NGC~4984 (\textit{top plot}, $i = 54^\circ$, $\text{PA} = -103^\circ$) and NGC~7755 (\textit{bottom plot}, $i = 53^\circ$, $\text{PA} = 94^\circ$).
For each plot, the \textit{left} panel shows the intensity map of the model. On \textit{right} panels we reproduce the maps of $h_4$ parameter from \citet{Gadotti_etal2020} with synthetic $h_4$ maps of our BL model shown on the inset plots. \textcolor{black}{The $h_4$ colourbar is the same as in Fig.~\ref{fig:2Dmaps_i0PA0}. }
}
\label{fig:real_gals_NGC4984}
\end{figure}

The most demanded in kinematic applications may be our BL model, although all our models show similar behaviour when the galaxy is inclined and rotated. We refer to the fact that all our models are galaxies with B/PS bulges. 
%Now we are wondering if based on the kinematics, we can
The question arises whether,  it is possible to say that the edge-on B/PS bulge in the BL model is identical to the face-on barlens based on the kinematics. For galaxies with $i>20^\circ$ and $\mathrm{PA}>45^\circ$, our models predict a ‘ring’ of negative values of $h_4$. The appearance of the ring in $h_4$ maps is consistent with the $h_4$ maps of the inclined galaxies NGC~4984 and NGC~7755 ($\mathrm{PA} \approx 90^\circ$) from \cite{Gadotti_etal2020}. As an illustration, we show the $h_4$ maps from \citet{Gadotti_etal2020} and model maps created from the BL model in Fig.~\ref{fig:real_gals_NGC4984}. The model was positioned in the same way relative to an observer as the galaxies in question.
\par
However, the presence of the ring of negative values of the parameter $h_4$ is not an exclusive feature of the BL model. With this position of the bar, the ring appears for all our models. The existence of such a ring is not a signature of a peanut. It is associated with the complicated contribution of other velocity components, $v_\mathrm{R}$ and $v_\varphi$, to the LOSVD, in addition to $v_z$. As we have shown above, a signature of the B/PS bulge through the $h_4$ parameter appears only along the major axis for low inclinations of the galaxy.
%%%%%%%%%%%%%%%%%%%%%%%%%%%%%%%%%%%%%%%%%%%%%%%%%%
%%%%%%%%%%%%%%%%%%%%%%%%%%%%%%%%%%%%%%%%%%%%%%%%%%

% Fig ??

\par
\begin{figure}
\centering
\includegraphics[width=\linewidth]{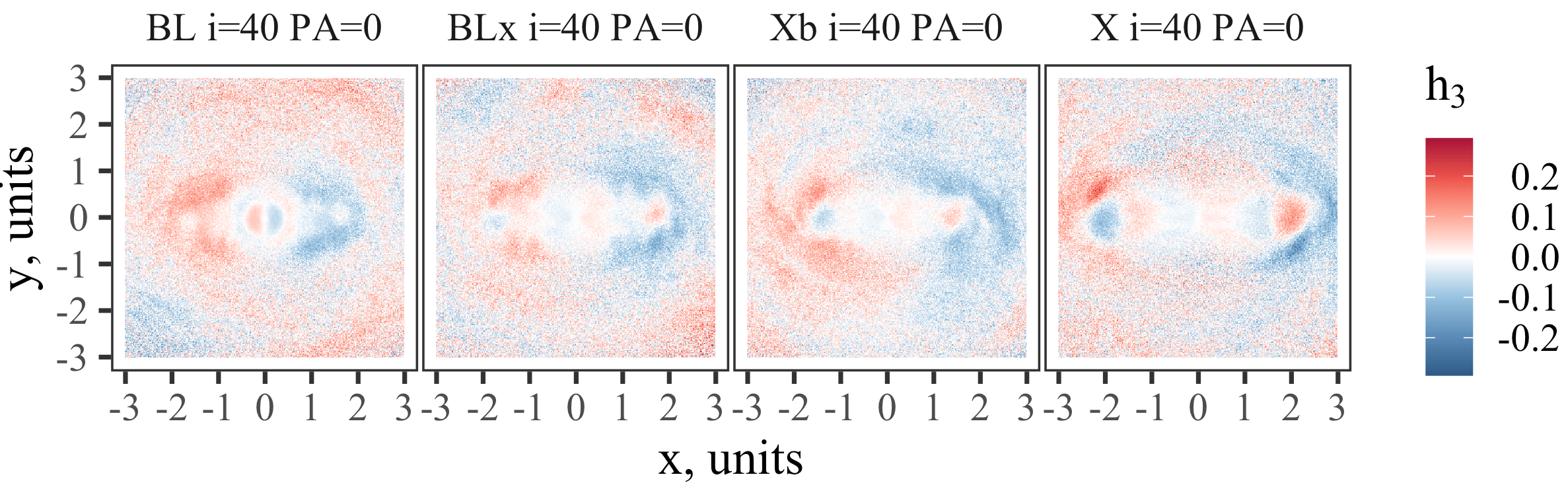}
\caption{The preliminary maps of $h_3$ parameter for all considered models at $\text{PA} = 0^\circ$, $i = 40^\circ$}
\label{fig:h3_maps}
\end{figure}

We stress that this work is primarily dedicated to the detection of the vertical density distribution features from the stellar kinematics. To make our point clear, we focus on the values of the $h_4$ parameter alone for now. However, the LOSVDs of the models contain a lot of information, for example, we do not use the `skewness' of the LOSVD~(the $h_3$ parameter). In Sec.~\ref{sec:LOSVD_i0} and \ref{sec:LOSVD_inclination_between_models}, we do find a minor difference in the $h_4$ maps depending on the model, but eventually conclude that these differences are insufficient to clearly distinguish them. This situation changes when one starts considering the $h_3$ parameter too. 
In Fig.~\ref{fig:h3_maps}, we provide preliminary $h_3$ maps for all our models, where one can see that the BL model has a distinctive feature in its central part~\textcolor{black}{(inside 0.5 units radius)}. This feature is associated with the rotation of the inner parts of the barlens but does not characterize its vertical structure. We tentatively connect this feature with \blu component, which is also responsible for the slightly elevated $h_4$ values in the same area (cf. the rightmost and the leftmost panels of the Fig.~\ref{fig:d4_exclude_orbits}), ultimately leaving the problem of the detection of kinematic characteristics of barlens galaxies using both $h_3$ and $h_4$ parameters to a future paper focused on a more detailed comparison of BL model kinematics with available observational data.

%%%%%%%%%%%%%%%%%%%%%%%%%%%%%%%%%%%%%%%%%%%%%%%%%%
%%%%%%%%%%%%%%%%%%%%%%%%%%%%%%%%%%%%%%%%%%%%%%%%%%
% Fig.~15

%%%%%%%%%%%%%%%%%%%%%%%%%%%%%%%%%%%%%%%%%%%%%%%%%%
%%%%%%%%%%%%%%%%%%%%%%%%%%%%%%%%%%%%%%%%%%%%%%%%%%

%%%%%%%%%%%%%%%%%%%%%%%%%%%%%%%%%%%%%%%%%%%%%%%%%%
\section{Conclusions}
\label{sec:conclusions}
% 7.
%%%%%%%%%%%%%%%%%%%%%%%%%%%%%%%%%%%%%%%%%%%%%%%%%%
%%%%%%%%%%%%%%%%%%%%%%%%%%%%%%%%%%%%%%%%%%%%%%%%%%
We have explored four $N$-body models that have a gradual progression in a face-on bar morphology and imitate real galaxies from galaxies with a barlens to featureless barred galaxies. 
Like our predecessors \citep{Debattista_etal2005,Iannuzzi_Athanassoula2015}, we focused on the analysis of 2D maps of the $h_4$ parameter to probe the presence of a ``peanut'' in cases where this morphological feature is not visually evident. These maps, produced on the basis of model data cubes with a spectral resolution close to that available, for example, for the data from the TIMER project \citep{Gadotti_etal2019}, can be compared with real observational data.
We have studied in detail the 2D kinematic maps for various orientations of the bar to the line of sight ($i=0^\circ-60^\circ$, $\mathrm{PA}=0^\circ - 90^\circ$). Our uniform grid of disc inclinations and position angles of the bar major axis covers almost all possible variations in the position of the galaxy, except for the edge-on view.
\par
Based on the spectral analysis of the main frequencies of the orbits in the bar, we have previously identified separate subsystems, or orbital groups, assembled into a bar \citep{Smirnov_etal2021}. The groups differ in structure, primarily vertical, which means that they can also differ kinematically. We have analysed the contributions of the LOSVDs of individual orbital components to the cumulative LOSVD in order to correctly interpret the LOSVD moment maps and judge which orbital groups make the main contribution to the negative values of the parameter $h_4$.
\par
The main conclusions of the work are as follows.
\begin{itemize}
\item[1.] The characteristic vertical density distribution attributed to the B/PS bulges subsists in models with and without barlens and can be detected using the parameter $h_4$.
\item[2.] The data analysed to reproduce the well-established face-on kinematic signatures of the presence of a ``peanut'' (B/PS bulge): at low inclinations ($i<20^\circ$) along \textit{the major axis of the bar}, two symmetrical negative minima of the $h_4$ parameter are observed. Unlike other negative pixels on the 2D maps of the $h_4$ parameter, the negative pixels of this parameter along the major axis of the bar really delineate the boundaries of the B/PS bulge.
\item[3.] There are no significant morphological differences in the maps of the $h_4$ parameter for the model with a barlens (BL model) and without it (X model) at all inclinations and $\mathrm{PA}>45^\circ$, when a ring of negative $h_4$ values begins to form. 
An interpretation of observational data at this viewing position of the galaxy only on the basis of these maps does not allow one to probe the vertical structure of a barlens if it is present and co-exists with a B/PS bulge within the same galaxy.
\item[4.] With an increased inclination, the diagnostics based on the $h_4$ parameter is ruined due to a break in the connection of the $h_4$ parameter with the features of the vertical density distribution. 
\item[5.] We clearly distinguish between the influence of two factors on the morphology of the LOSVD moment maps. The first is the effect of the inclination (contamination of the LOSVD with velocity projections other than $v_z$). The second is the contamination of the cumulative LOSVD by the LOSVDs of different orbital subsystems or with different kinematics (fast and slow rotating components).
\item[6.]  The wide ring of negative values of $h_4$ is observed on 2D model maps at intermediate inclinations ($i=20^\circ-60^\circ$) and $\mathrm{PA}>45^\circ$. Such rings can be noticed on the $h_4$ maps for the barlens galaxies (e.g., NGC~3351, NGC~4984, NGC~7755, \citealp{Gadotti_etal2020}). However, rings are present in all our models, and their appearance can not be a distinguishing feature of the barlens model.
\item[7.] We analysed the impact of different orbital groups in the bar on the kinematic signatures of a peanut for all our models. For our models, different orbital groups make different contributions to the minima of $h_4$ on both sides along the major axis of the bar. For the barlens model, the minima of $h_4$ are associated with the x1 family of orbits, the main building material of a long narrow bar. If this orbital family is well represented in the bar, then it is this family that will give two-sided negative minima of $h_4$ along the major axis of the bar and signals about the features of the vertical density distribution~(B/PS bulge). Thus, the orbits supporting the barlens are not responsible for the kinematic signatures of the peanut.
\end{itemize}
%%%%%%%%%%%%%%%%%%%%%%%%%%%%%%%%%%%%%%%%%%%%%%%%%%
%%%%%%%%%%%%%%%%%%%%%%%%%%%%%%%%%%%%%%%%%%%%%%%%%%

%%%%%%%%%%%%%%%%%%%%%%%%%%%%%%%%%%%%%%%%%%%%%%%%%%
%%%%%%%%%%%%%%%%%%%%%%%%%%%%%%%%%%%%%%%%%%%%%%%%%%
\section*{Acknowledgements}
%%%%%%%%%%%%%%%%%%%%%%%%%%%%%%%%%%%%%%%%%%%%%%%%%%
%%%%%%%%%%%%%%%%%%%%%%%%%%%%%%%%%%%%%%%%%%%%%%%%%%
The authors express gratitude for the financial support from the Russian Science Foundation (grant no. 22-22-00376). \textcolor{black}{We are grateful to the anonymous reviewer for a careful and thorough reading of the manuscript.} We are also grateful to the TIMER team \citep{Gadotti_etal2019} for making their data publicly available. This work made use of Astropy:\footnote{\url{http://www.astropy.org}} a community-developed core Python package and an ecosystem of tools and resources for astronomy \citep{astropy:2013, astropy:2018, astropy:2022}, as well as NEMO stellar dynamics toolbox \citep{Teuben_1995}. The authors also thank Roberto Saglia for reading a preliminary version of this manuscript and a fruitful discussion regarding the optimal way to present the results of this work.
%%%%%%%%%%%%%%%%%%%%%%%%%%%%%%%%%%%%%%%%%%%%%%%%%%
%%%%%%%%%%%%%%%%%%%%%%%%%%%%%%%%%%%%%%%%%%%%%%%%%%
\section{Data availability}
The data underlying this article be shared on reasonable request to the corresponding author.

\bibliographystyle{mnras}
\bibliography{main} % if your bibtex file is called example.bib

%%%%%%%%%%%%%%%%%%%%%%%%%%%%%%%%%%%%%%%%%%%%%%%%%%
%%%%%%%%%%%%%%%%% APPENDICES %%%%%%%%%%%%%%%%%%%%%
\appendix
%%%%%%%%%%%%%%%%%%%%%%%%%%%%%%%%%%%%%%%%%%%%%%%%%%
%%%%%%%%%%%%%%%%%%%%%%%%%%%%%%%%%%%%%%%%%%%%%%%%%%
%%%%%%%%%%%%%%%%%%%%%%%%%%%%%%%%%%%%%%%%%%%%%%%%%%
\section{Effects of inclination: $h_4$ maps}
\label{app:inclination_full_maps}
\begin{figure}
\centering
\includegraphics[width=\linewidth]{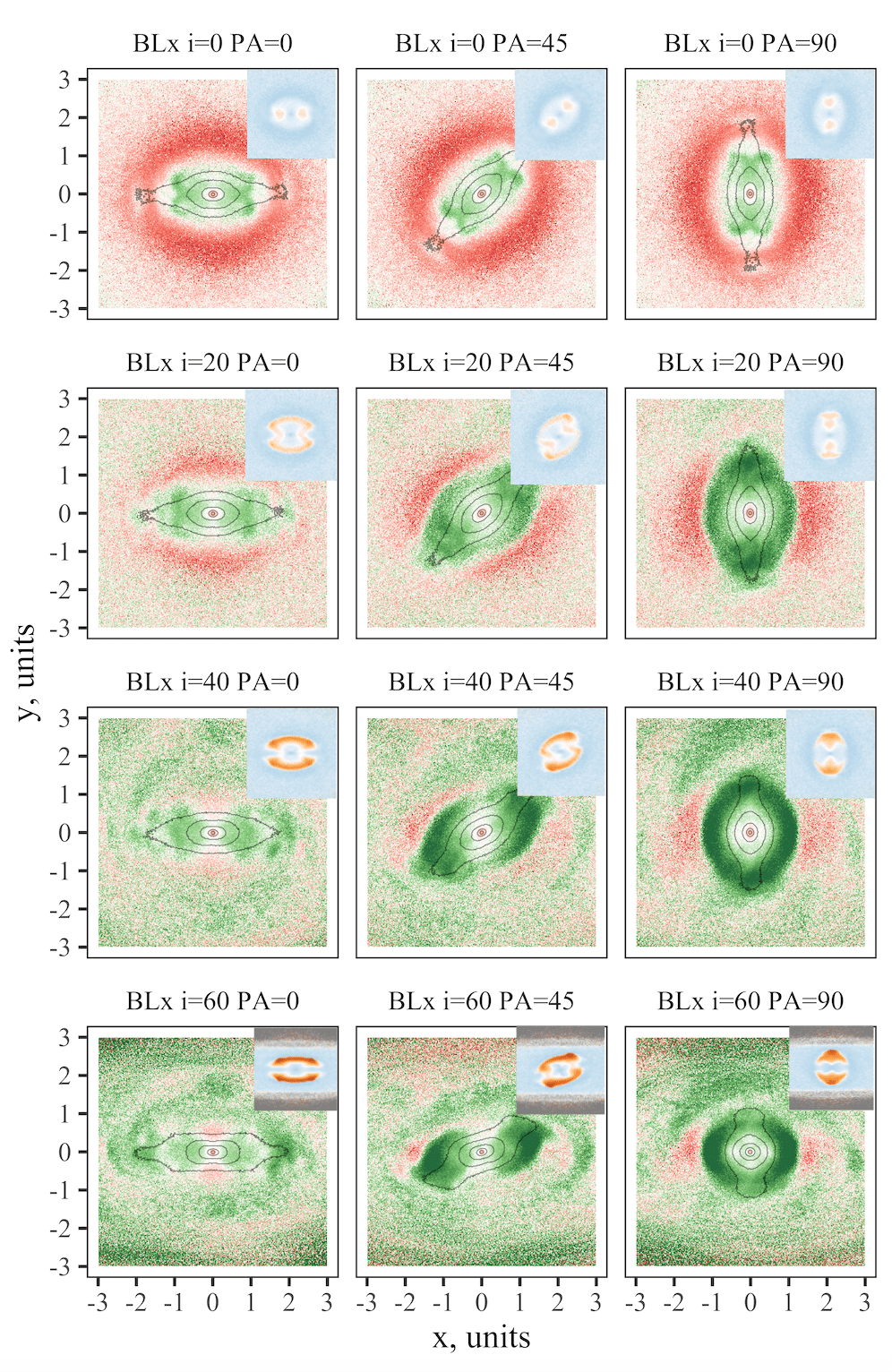}
\caption{Effects of the disc inclination ($i$) and PA in the $h_4$ maps for the BLx model. See Fig.~\ref{fig:BL_all_i} for all details.}%
\label{fig:BLx_all_i}
\end{figure}

\begin{figure}
\centering
\includegraphics[width=\linewidth]{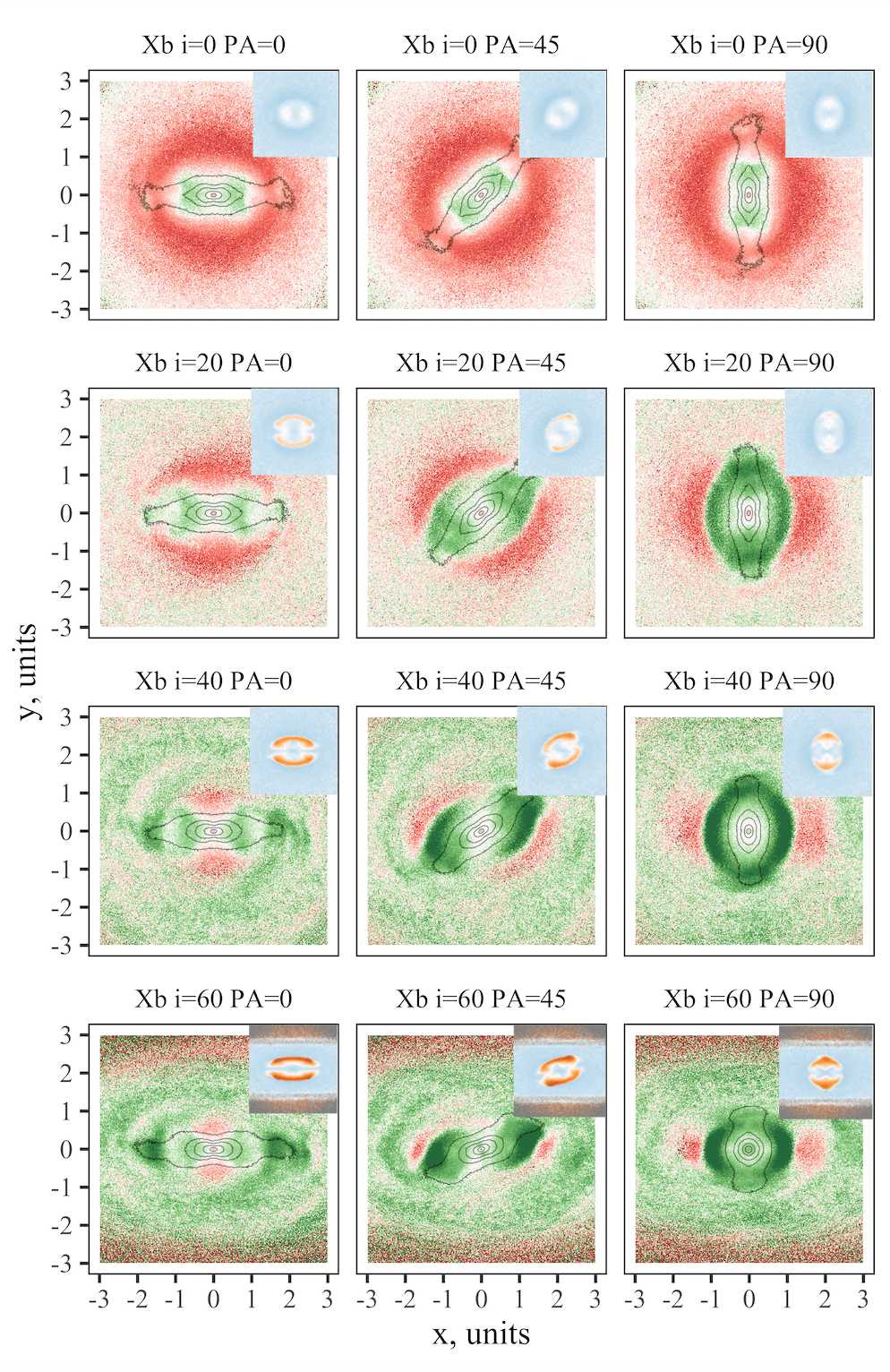}
\caption{Effects of the disc inclination ($i$) and PA in the $h_4$ maps for the Xb model. See Fig.~\ref{fig:BL_all_i} for all details.
}%
\label{fig:Xb_all_i}
\end{figure}

\begin{figure}
\centering
\includegraphics[width=\linewidth]{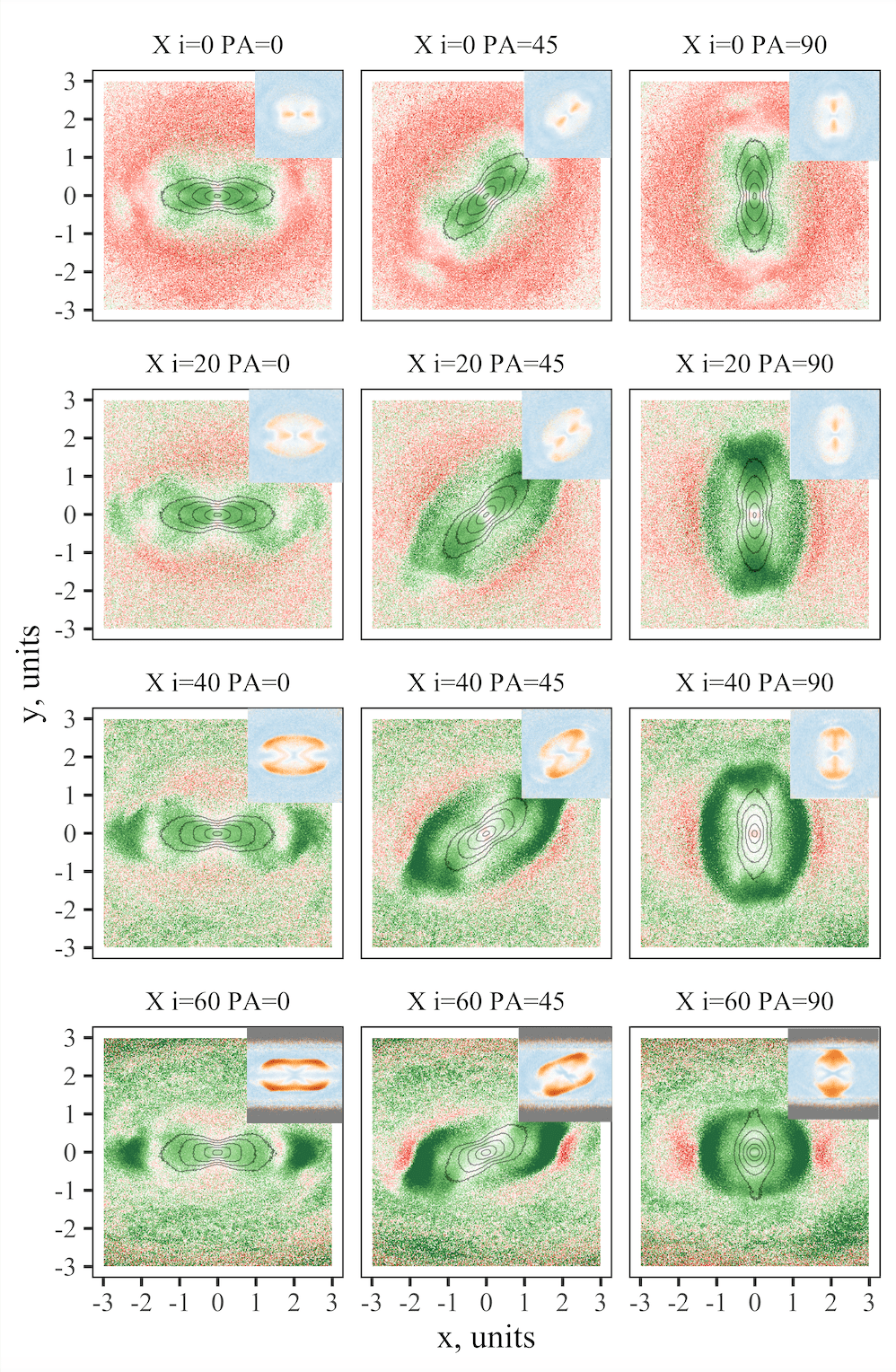}
\caption{Effects of the disc inclination ($i$) and PA in the $h_4$ maps for the X model. See Fig.~\ref{fig:BL_all_i} for all details.
}%
\label{fig:X_all_i}
\end{figure}

%%%%%%%%%%%%%%%%%%%%%%%%%%%%%%%%%%%%%%%%%%%%%%%%%%
%%%%%%%%%%%%%%%%%%%%%%%%%%%%%%%%%%%%%%%%%%%%%%%%%%
In Section~\ref{sec:LOSVD_inclination_bl}, we provide the $h_4$ maps for different inclinations and position angles for the BL model. 
Figs.~\ref{fig:BLx_all_i}, \ref{fig:Xb_all_i}, \ref{fig:X_all_i} show the maps similar to those presented in Fig.~\ref{fig:BL_all_i}, for the BLx, Xb, and X models. Despite the differences between $h_4$ maps for a face-on view, which depend on the exact bar morphology, the inclination effects are similar and lead to the same kinematic features for all presented B/PS bulges. As was shown in the Sections~\ref{sec:LOSVD_inclination_bl}, ~\ref{sec:LOSVD_inclination_between_models}, ~\ref{sec:orbital_groups}, ~\ref{sec:orbit_exl_4090} all B/PS bulges form the rings of $h_4$ minima for $i>20^\circ$ and $\mathrm{PA}>45^\circ$. We also note that the size of the ring depends on the exact size of the B/PS bulge.

%%%%%%%%%%%%%%%%%%%%%%%%%%%%%%%%%%%%%%%%%%%%%%%%%%
%%%%%%%%%%%%%%%%%%%%%%%%%%%%%%%%%%%%%%%%%%%%%%%%%%
%%%%%%%%%%%%%%%%%%%%%%%%%%%%%%%%%%%%%%%%%%%%%%%%%%
%%%%%%%%%%%%%%%%%%%%%%%%%%%%%%%%%%%%%%%%%%%%%%%%%%

%%%%%%%%%%%%%%%%%%%%%%%%%%%%%%%%%%%%%%%%%%%%%%%%%%
%%%%%%%%%%%%%%%%%%%%%%%%%%%%%%%%%%%%%%%%%%%%%%%%%%
%%%%%%%%%%%%%%%%%%%%%%%%%%%%%%%%%%%%%%%%%%%%%%%%%%
%%%%%%%%%%%%%%%%%%%%%%%%%%%%%%%%%%%%%%%%%%%%%%%%%%

%%%%%%%%%%%%%%%%%%%%%%%%%%%%%%%%%%%%%%%%%%%%%%%%%%
%%%%%%%%%%%%%%%%%%%%%%%%%%%%%%%%%%%%%%%%%%%%%%%%%%
%%%%%%%%%%%%%%%%%%%%%%%%%%%%%%%%%%%%%%%%%%%%%%%%%%
%%%%%%%%%%%%%%%%%%%%%%%%%%%%%%%%%%%%%%%%%%%%%%%%%%

%%%%%%%%%%%%%%%%%%%%%%%%%%%%%%%%%%%%%%%%%%%%%%%%%%
%%%%%%%%%%%%%%%%%%%%%%%%%%%%%%%%%%%%%%%%%%%%%%%%%%
\section{Dissection of the bar into orbital groups for BLx and Xb model}
\label{app:blx_xb_orbits}
\begin{figure*}
\centering
\includegraphics[width=1\linewidth]{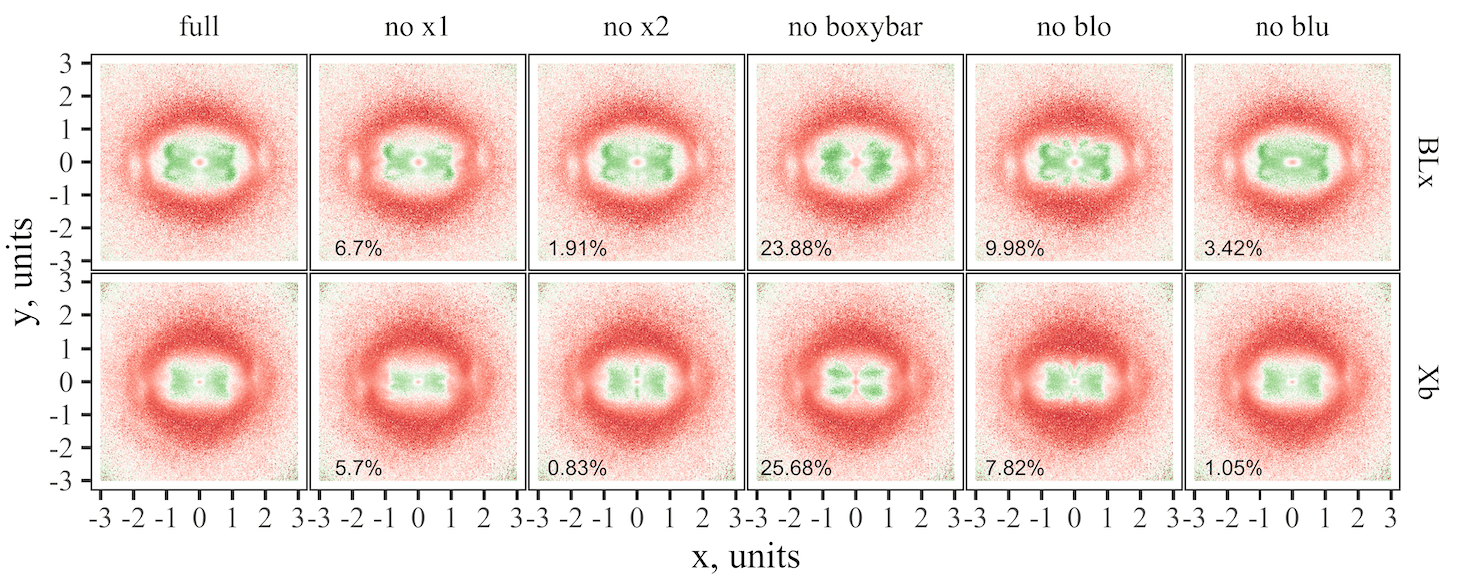}
\caption{The maps of $h_4$ for the BLx (\emph{top} row) and Xb (\emph{bottom} row) models at face-on position with orbital groups excluded in turn.  See Fig.~\ref{fig:d4_exclude_orbits} for all details.}
\label{app:d4_exclude_orbits_blx_xb}
\end{figure*}

\begin{figure*}
\centering
\includegraphics[width=1\linewidth]{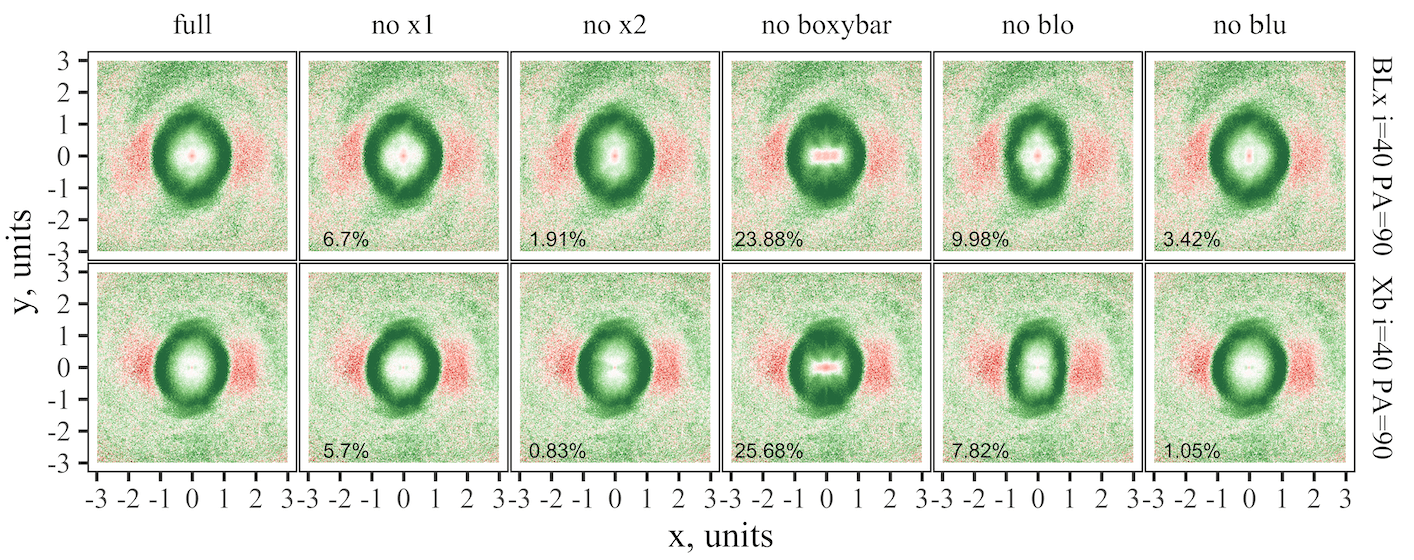}
\caption{The maps of $h_4$ for the BLx (\emph{top} row) and Xb (\emph{bottom} row) models at $i=40^\circ$, $\mathrm{PA}=90^\circ$ with orbital groups excluded in turn.  See Fig.~\ref{fig:h4_exclude_orbits_4090} for all details.}  
\label{app:h4_exclude_orbits_4090}
\end{figure*}

%%%%%%%%%%%%%%%%%%%%%%%%%%%%%%%%%%%%%%%%%%%%%%%%%%
%%%%%%%%%%%%%%%%%%%%%%%%%%%%%%%%%%%%%%%%%%%%%%%%%%
Section~\ref{sec:orbital_groups} contains the $h_4$ maps for the BL and X models. Here we provide the same maps for the BLx and Xb models. Both of them mainly consist of a quarter of boxy orbits, which contribute to the $h_4$ maps in the face-on case~(Fig.~\ref{app:d4_exclude_orbits_blx_xb}). However this contribution is hardly to notice. We refer it to the smearing of the peanut signatures by the classical bulge in the centre (for the BL model the classical bulge is compact, the X model has no a classical bulge). Fig.~\ref{app:h4_exclude_orbits_4090} shows that the exclusion of any of orbital group of the bar does not change the $h_4$ maps for the inclined galaxies. 

%%%%%%%%%%%%%%%%%%%%%%%%%%%%%%%%%%%%%%%%%%%%%%%%%%
%%%%%%%%%%%%%%%%%%%%%%%%%%%%%%%%%%%%%%%%%%%%%%%%%%
%%%%%%%%%%%%%%%%%%%%%%%%%%%%%%%%%%%%%%%%%%%%%%%%%%
%%%%%%%%%%%%%%%%%%%%%%%%%%%%%%%%%%%%%%%%%%%%%%%%%%

%%%%%%%%%%%%%%%%%%%%%%%%%%%%%%%%%%%%%%%%%%%%%%%%%%
%%%%%%%%%%%%%%%%%%%%%%%%%%%%%%%%%%%%%%%%%%%%%%%%%%
%%%%%%%%%%%%%%%%%%%%%%%%%%%%%%%%%%%%%%%%%%%%%%%%%%
%%%%%%%%%%%%%%%%%%%%%%%%%%%%%%%%%%%%%%%%%%%%%%%%%%

%%%%%%%%%%%%%%%%%%%%%%%%%%%%%%%%%%%%%%%%%%%%%%%%%%
% Don't change these lines
\bsp	% typesetting comment
%%%%%%%%%%%%%%%%%%%%%%%%%%%%%%%%%%%%%%%%%%%%%%%%%%
%%%%%%%%%%%%%%%%%%%%%%%%%%%%%%%%%%%%%%%%%%%%%%%%%%
\label{lastpage}
\end{document}